\documentclass[11pt]{article}
\usepackage{natbib}
\usepackage{amsmath,amsfonts,amsthm}
\usepackage{bm}
\usepackage{color}
\usepackage{hyperref}
\hypersetup{colorlinks=true,
            linkcolor={red},
            citecolor={blue}}
\theoremstyle{definition}
\newtheorem{assumption}{Assumption}
\usepackage{amsmath,amssymb,fullpage}
\usepackage{enumerate}
\usepackage{graphicx}
\usepackage{makeidx}
\usepackage[utf8]{inputenc}
\usepackage{wrapfig}
\usepackage{mathrsfs}
\usepackage{amssymb}
\usepackage{latexsym}
\usepackage{amsbsy}
\usepackage{color}
\usepackage{bbm}
\usepackage{mathrsfs}
\usepackage{todonotes}
\usepackage{url}

\usepackage[ruled, lined, linesnumbered, commentsnumbered, longend]{algorithm2e}
\usepackage{xcolor}

\SetCommentSty{mycommfont}

\usepackage{wasysym}
\def\email#1{\it #1\par}

\providecommand{\otherindexspace}[1]{}

\usepackage{amsthm}
\newtheorem{theorem}{Theorem}[section]
\newtheorem{lemma}[theorem]{Lemma}
\newtheorem{proposition}[theorem]{Proposition}
\newtheorem{remark}[theorem]{Remark}
\newtheorem{definition}[theorem]{Definition}

\newtheorem{conjecture}[theorem]{Conjecture}
\newtheorem{corollary}[theorem]{Corollary}

\newtheorem{example}[theorem]{Example}

\definecolor{falured}{rgb}{0.5, 0.09, 0.09}

\usepackage{comment}

\usepackage{dsfont}

\providecommand{\keywords}[1]
{
  \small	
  \textbf{\textit{Keywords---}} #1
}

\newcommand{\indep}{\perp \!\!\! \perp}

\numberwithin{equation}{section}

\title{A fixed point approach for computing actuarially fair Pareto optimal risk-sharing rules}
\author{Fallou NIAKH \thanks{Centre de Recherche en Économie et de Statistiques (CREST), CNRS, École Polytechnique, GENES, ENSAE Paris, Institut Polytechnique de Paris, 91120 Palaiseau, France, Email: \email fallou.niakh@ensae.fr}}

\date{}

\begin{document}
\maketitle

\begin{abstract}
    Risk-sharing is one way to pool risks without the need for a third party. To ensure the attractiveness of such a system, the rule should be accepted and understood by all participants. A desirable risk-sharing rule should fulfill actuarial fairness and Pareto optimality while being easy to compute. This paper establishes a one-to-one correspondence between an actuarially fair Pareto optimal (AFPO) risk-sharing rule and a fixed point of a specific function. A fast numerical method for computing these risk-sharing rules is also derived. As a result, we are able to compute AFPO risk-sharing rules for a large number of heterogeneous participants in this framework.  
\end{abstract}

\section{Introduction}
Risk-sharing has received considerable attention as a mechanism that allows members of a community to combine their resources for the benefit of all. It can be used as an alternative to transferring risk to an insurance company, to a financial market, among others. 

In a risk-sharing scheme, participants agree to pool together their risk-bearing capacity and share losses among themselves based on some rule \emph{agreed upon prior} to the observation of any loss. The main challenge in this area is to determine this \emph{ex-ante} rule that has to be accepted by all participants. When individual losses of participants are identically distributed and they have identical risk aversions, a natural risk-sharing rule involves allocating an equal share of the aggregate risk to each participant, resulting in a uniform distribution of risk across all pool members. Apart from this case, the problem of risk-sharing becomes more complex. 

Several papers, including \cite{buhlmann_optimal_1979}, \cite{denuit_convex_2012}, \cite{schumacher_linear_2018}, and \cite{abdikerimova_peer--peer_2019}, proposed different risk-sharing mechanisms. \emph{Linear} risk-sharing rules are often used according to \cite{schumacher_linear_2018}. In that case, participants agree to take a fixed percentage of the total loss in accordance with the expected values of the risks they bring to the pool compared to the total expected loss. Nevertheless, \emph{linear} risk-sharing rules do not take into account the variability of individual losses and risk aversions. To address this, \emph{non-linear} risk-sharing rules have been investigated. A recent \emph{non-linear} risk-sharing rule is the conditional mean risk-sharing rule proposed in \cite{denuit_convex_2012} where each participant contributes the conditional expectation of the loss brought to the pool, given the total loss experienced by the entire pool (\cite{denuit_risk_2021}). Some attractive properties of the conditional mean risk allocation have been studied in \cite{denuit_size-biased_2019, denuit_large-loss_2020, denuit_risk-sharing_2022, jiao_axiomatic_2022}. 

Two important considerations that arise when designing such risk-sharing rules are \emph{actuarial fairness} and \emph{Pareto optimality}. \emph{Actuarial fairness} refers to the fact that the expected loss for each participant before and after risk-sharing remains the same. A \emph{Pareto optimal} risk-sharing rule is such that there does not exist another risk-sharing rule which leads to a better financial situation for all participants and a strictly better financial situation for at least one participant. Here, we investigate risk-sharing rules that are \emph{actuarially fair Pareto optimal (AFPO)}.

In this paper, a convenient representation of AFPO risk-sharing rules is provided. More precisely, we establish a one-to-one correspondence between an \emph{AFPO} risk-sharing rule and a fixed point of a specific function. Fast numerical methods to compute AFPO risk-sharing rules are essential for applications. The model used as a basis for risk-sharing is a one-period model with a finite number of von Neumann–Morgenstern participants. As inputs to the design problem, we consider participants’ preferences, specified by disutility functions, and the joint distribution of individual risks which they bring into the pool.

The fixed-point approach has been extensively used for solving a variety of problems in different fields. This strategy provides a powerful tool for showing the existence and uniqueness of solutions under weak assumptions and lends itself well to numerical implementation. For example, this method is well-known in the field of economics to find the general equilibrium of supply and demand in a market: the prices and quantities in the market adjust iteratively until a stable solution is reached where demand equals supply (see e.g. \cite{herings_equilibrium_2002, herings_pareto_2005}). In the field of differential equations, the fixed-point approach is commonly used to establish the existence and uniqueness of solutions for Backward Stochastic Differential Equations (BSDEs) and to compute them numerically (see e.g. \cite{carmona_lectures_2016}). Recently, in the field of artificial intelligence, this approach has gained prominence in reinforcement learning algorithms. These algorithms use an iterative process where the agent's actions are updated based on feedback and rewards until an optimal policy is determined.

In the field of risk sharing, which is the subject of our study, this fixed-point approach has been first investigated by \cite{pazdera_composite_2017}. They reformulate the problem of finding risk-sharing rules that are both financially fair and Pareto-efficient as a fixed-point problem. 
In their framework, it is assumed that a global financial risk of the market is to be shared. State prices, which are used to determine financial fairness, are given. \cite{pazdera_composite_2017} study the problem of finding an efficient and fair ex-ante rule for division of an uncertain monetary outcome. They show that efficient and financially fair allocation rules are in one-to-one correspondence with positive eigenvectors of a nonlinear homogeneous and monotone mapping associated to the risk-sharing problem. By establishing relevant properties of this mapping, they prove the existence and uniqueness of solutions via nonlinear Perron–Frobenius theory, as well as the global convergence of the associated iterative algorithm. We follow the work of \cite{pazdera_composite_2017} in risk-sharing specifically in the context of insurance. However, this requires theoretical and algorithmic adaptations to address the unique characteristics and requirements of the insurance industry.

Unlike \cite{pazdera_composite_2017}, where a global financial risk of the market is shared between participants, in our insurance case, this risk is dependent on the vector of individual random losses. Also, to establish our actuarial fairness condition, we do not require the existence of a risk-neutral probability. Another difference lies in the desired properties due to specific insurance characteristics concerning risk-sharing rules. Indeed, in \cite{pazdera_composite_2017}, they assumed that the risk-sharing rules are real-valued and that agents can choose the domain to which these rules belong. In our insurance framework, however, the participants' risk-sharing rules should be non-negative and bounded by their individual maximum losses. Therefore, no specific choices or preferences are attributed to these risk-sharing rules by the participants themselves. Finally, the risk-sharing rules should be zero if nobody makes a claim., i.e. these rules vanish at 0.

This paper provides several notable significant contributions to the current literature on risk-sharing.

\begin{enumerate}
    \item By using the one-to-one correspondence between AFPO risk-sharing rules and fixed points of a specific function, we have established the existence and uniqueness of AFPO risk-sharing rules in this framework.
    \item A fast numerical algorithm is provided to compute AFPO risk-sharing rules. The convergence of this algorithm is proved.
    \item Our approach can be used for a large number of heterogeneous participants, such as more than 1000 participants, for example.
    \item We provide several numerical illustrations of the fixed point algorithm.
\end{enumerate}

The paper is organized as follows. Section \ref{sect2} introduces the risk-sharing framework. In Section \ref{sect3}, we present the fixed point approach for determining AFPO risk-sharing rules. Numerical examples that illustrate the efficiency of our algorithm are provided in Section \ref{sect4}. Section \ref{cxd} discusses the comparability of AFPO risk-sharing rules in the convex order sense. Section \ref{sect5} concludes.

\section{Risk-sharing framework \label{sect2}}

\subsection{Notation and assumptions }
We consider a risk-sharing pool composed of $n$ participants, numbered $i \ \in \{1, \dots, n \}$. Each of them faces a risk denoted by the random variable $X_i$. By risk, we mean a non-negative random variable representing monetary losses caused by some insurable peril over one period (a calendar year, say). Throughout the paper, one assumes that $X_1,...,X_n$ are valued in $[0, \infty)$. The case where the risk brought by one participant to the pool is not random, but equal to some non-negative constant $c$ with probability $1$ is not allowed in this framework. These losses are defined on a common probability space $(\Omega, \mathcal{F}, P)$. This space is assumed to be rich enough to contain all the random variables mentioned throughout the paper. The marginal distribution functions of the individual losses are denoted by $F_1, ..., F_n$, respectively. A $n$-dimensional random
vector of losses $\textbf{X} = (X_1, \dots ,X_n)$ will be called a pool (of losses). Then the entire pool $\textbf{X}$ has an aggregate risk given by the sum $S=\sum_{i=1}^{n}X_i$.\bigskip

Participants make decisions in an uncertain environment, where each of them ranks a risky outcome according to their expected utility. Utility functions are the natural choice when dealing with gains. However, in this paper, one will work with non-negative losses. Therefore, one can simplify by switching to disutility functions $v_i$ (as in \cite{buhlmann_optimal_1979}) defined as:
\begin{equation}
    v_i(s)=-u_i(-s)
\end{equation}
where $u_i$ is participant $i$'s utility function for $i \ \in \{1, \dots, n \}$.
The risk tolerance (i.e. the inverse of the Arrow-Pratt coefficient of absolute risk aversion) for participant $i$ is defined as follows:
\begin{equation}
    T_i(s)=\frac{v_i^{\prime}(s)}{v_i^{\prime\prime}(s)}.
    \label{risktol}
\end{equation}
Define the largest value for the loss $X_i$ as
$$\max[X_i]= \sup \ \{x \ \in \ \mathbb{R} \ | \ F_i(x)<1 \}.$$
It seems to be unfair to ask participants to contribute more than their maximal loss value $\max[X_i]$. This is in essence the \emph{no-ripoff} requirement (see \cite{denuit_risk-sharing_2022} for more details). As a result, the participants' disutility functions are taken to be defined on $[0, \max[X_i])$ with $\max[X_i]  \in  (0, \infty]$, and will always be assumed to satisfy Assumption \ref{H1}.
\begin{assumption}
For each $i \ \in \{1, \dots, n \}$, the function $v_i: [0, \max[X_i]) \rightarrow \mathbb{R}$ is twice continuously differentiable, increasing, and strictly convex. Moreover, the following Inada conditions are satisfied:
\begin{equation}
  \lim\limits_{s  \rightarrow 0} v_{i}^{\prime}(s)=0, \ \ \ \lim\limits_{s \rightarrow \max[X_i]} v_{i}^{\prime}(s)=\infty.
  \label{formule22}
\end{equation}
\label{H1}
\end{assumption}
\noindent Based on this assumption, all marginal disutilities $v_i^{\prime}$ are continuous and increasing functions whose range spans the non-negative real axis. The inverse marginal disutility of participant $i$ will be denoted by $I_i$. In other words, $I_i$ is the function from $[0, \infty)$ to $[0, \max[X_i])$ that is defined implicitly by
\begin{equation}
    v_{i}^{\prime}\left(I_{i}(z)\right)=z \ \forall \ z \geq 0.
    \label{formule23}
\end{equation}
The inverse marginal disutility is a continuous and increasing function that has the interval $[0, \max[X_i])$ as its image.
Note the fact that the image of $I_i$ is $[0, \max[X_i])$ will be very important for the following. Indeed, it will ensure that risk-sharing rules will always be non-negative and each will not exceed the maximum loss of the associated participant. 
Let us define the support of the total risk $S$ as follows
\begin{equation}
     A:=\left[0, \sum_{i=1}^{n} \max[X_i]\right].
    \label{formule24}
\end{equation}

\begin{assumption}
$$\sum_{i=1}^{n} \max[X_i] < \infty .$$
\label{ass2}
\end{assumption}
\noindent Assumption \ref{ass2} implies that the losses $X_1, \dots, X_n$ are bounded. The space of functions from $A$ to $\mathbb{R}_{+}$ is denoted by $\mathcal{L}(A, \mathbb{R}_{+})$.

\subsection{Definitions}

\begin{definition}[Risk-sharing rule]
\emph{A risk-sharing rule $\mathbf{h}$ is a mapping which transforms any pool $\textbf{X} = (X_1, X_2, \dots,X_n)$ into another random vector $\mathbf{h}(\textbf{X})=\left(h_1(\textbf{X}), h_2(\textbf{X}), \dots, h_n(\textbf{X})\right)$ where the (measurable) functions $h_i : \left[0, \max[X_1]\right] \times \dots \times \left[0, \max[X_n]\right] \to \mathbb{R}_+$ are such that
\begin{equation}
    \sum_{i=1}^{n} h_{i}(\textbf{X})=S.
    \label{self}
\end{equation}
}
\end{definition}

\begin{remark}
At the end of the period, each agent will pay $\left(h_1(\textbf{x}), h_2(\textbf{x}), \dots, h_n(\textbf{x})\right)$ where $\textbf{x}$ is the observed realization of $\textbf{X}$. Condition \eqref{self} is called the \emph{full risk allocation} condition in \cite{denuit_risk-sharing_2022}, or also \emph{top-down consistency} in \cite{guan_impossibility_2022}.
\end{remark}
\noindent An important subclass of risk-sharing rules consists of 

$$\left(h_1(\textbf{X}), h_2(\textbf{X}), \dots, h_n(\textbf{X})\right)=\left(g_1(S), g_2(S), \dots, g_n(S)\right)$$
for some functions $g_1, \dots, g_n$ in $\mathcal{L}(A, \mathbb{R}_{+})$. A risk-sharing rule fulfilling this property is called \emph{aggregate risk-sharing rule} by \cite{denuit_risk-sharing_2022} and \emph{'Non-olet' risk-sharing rule} by \cite{feng_unified_2022}. In the sequel, we will focus exclusively on \emph{aggregate} or \emph{'non-olet'} risk-sharing rules. Indeed, \cite{borch_equilibrium_1962} established that under mild assumptions, participants' optimal risk-sharing depends only on the aggregate loss $S$.\bigskip

Here, AFPO risk-sharing rules are studied. 

\begin{definition}[actuarially fair Pareto optimal risk-sharing rules]
\mbox{}
    \begin{itemize}
        \item \emph{A 'non-olet' risk-sharing rule $\mathbf{h}$ for the given total risk $S$ is \textbf{actuarially fair} if participants do neither gain nor lose from risk-sharing, in the sense that  their expected contribution (by joining the pool) is equal to their expected loss (when staying alone), i.e.,}

            \begin{equation}
                E\left[h_{i}(S)\right]=E[X_i] \quad\forall \ i \ \in \{1, \dots, n \}.
                \label{formule32}
            \end{equation}
        \item \emph{ A 'non-olet' risk-sharing rule $\left(h_{1}, \ldots, h_{n}\right)$ is\textbf{ Pareto optimal} for $S$ if there does not exist a risk-sharing rule $\left(\tilde{h}_{1}, \ldots, \tilde{h}_{n}\right)$ such that $$\mathbb{E}[v_{i}(\tilde{h}_{i}(S))] \leq \mathbb{E}[v_{i}(h_{i}(S))] \ \forall \ i \ \text{and} \ \exists \ j \ \text{s.t.} \ \mathbb{E}[v_{j}(\tilde{h}_{j}(S))] < \mathbb{E}[v_{j}(h_{j}(S))].$$ }
    \end{itemize}
\noindent \emph{Actuarially fair Pareto optimal (AFPO) risk-sharing rules are Pareto optimal risk-sharing rules under the condition of actuarial fairness.}
\label{def23}
\end{definition}

\begin{remark}
Definition \ref{def23} is in line with \cite{buhlmann_optimal_1979}. Notice that the definition of Pareto-optimality given above is not the only one possible. For instance, \cite{denuit_convex_2012} require that there is no risk-sharing rule $\left(\tilde{h}_{1}, \ldots, \tilde{h}_{n}\right)$ for $S$ such that the following relation holds
$$\tilde{h}_i(S) \preceq_{\mathrm{CX}} h_i(S), \quad\forall \ i \ \in \{1, \dots, n \}$$
with at least one strict inequality, where $\preceq_{\mathrm{CX}}$ refers to the \emph{convex order}. For a thorough description of the convex order and its applications in insurance studies, we refer the reader to \cite{denuit_actuarial_2005}. It is important to note that the Pareto optimality definition used here is not stronger or weaker than the one based on the convex order.
\end{remark}

\subsection{Representation of AFPO risk-sharing rules}
\cite{borch_equilibrium_1962} observed that Pareto optimality could be obtained for every outcome of $S$, and therefore, did not depend upon the distribution of $S$. \cite{eeckhoudt_economic_2005} (Chapter 10) provides more details on Pareto optimal risk-sharing rules. \cite{borch_equilibrium_1962} characterized these solutions as follows:
\begin{theorem}
\emph{
A risk-sharing rule $\left(h_{1}, \ldots, h_{n}\right)$ is Pareto optimal for any given total risk $S$ taking values in the domain $A$ if and only if there exist a function $J: A \rightarrow \mathbb{R}_{+}$ and positive constants $\alpha_{1}, \ldots, \alpha_{n}$ such that
\begin{equation}
    \alpha_{i} v_{i}^{\prime}\left(h_{i}(s)\right)=J(s)
    \label{formule33}
\end{equation}
for all $s \in A$ and for all $i=1, \ldots, n$.}
\label{thm1}
\end{theorem} 
 Borch's characterization arises from the problem of minimizing the global disutility of participants under the feasibility constraint given in \eqref{self}. The mapping $J$ is the Lagrange multiplier associated with this constraint. We refer to \cite{eeckhoudt_economic_2005} (Section 10.3) for a thorough description of this minimization problem and derivation of equation \eqref{formule33}. The expression of the risk-sharing rules in \eqref{formule33} can be characterized by the mapping $J$ and the weights $\alpha_1,...,\alpha_n$ using the inverse marginal disutilities (cf. \eqref{formule23}) as follows:

\begin{equation}
    h_{i}(s)=I_{i}\left(J(s) / \alpha_{i}\right).
    \label{formule34}
\end{equation}
Thus, the feasibility condition \eqref{self} can be rewritten by replacing the risk-sharing rules with their new expressions above. In other words, for all $s \in A$ the following new feasibility condition holds:

\begin{equation}
    \sum_{i=1}^{n} I_{i}\left(J(s) / \alpha_{i}\right)=s.
    \label{formule35}
\end{equation}
One can notice that if $\boldsymbol{\alpha}=(\alpha_1,...,\alpha_n)$ and $s$ are known then equation \eqref{formule35} allows us to determine $J(s)$ in a unique way. Indeed, the function  $z \longmapsto \sum_{i=1}^{n} I_{i}\left(z / \alpha_{i}\right)$ is increasing. We can therefore write $J$ as a function of $s$ and $\boldsymbol{\alpha}$: $J(s; \boldsymbol{\alpha})$. This means that Pareto optimal risk-sharing rules are \emph{fully parameterized} by disutility weights $\alpha_1,...,\alpha_n$. In other words, for each positive vector $\boldsymbol{\alpha}$, there exists a Pareto optimal risk-sharing rule associated with it. Thus, Borch's Theorem \ref{thm1} leads to an uncountable set of risk-sharing rules that do not entirely describe the "best" risk-sharing schemes. Therefore, some authors (see \cite{buhlmann_optimal_1979}) have proposed versions of this theorem where Pareto optimal risk-sharing rules are \emph{constrained}. \\
The actuarial fairness condition \eqref{formule32} can be combined with Borch's Pareto optimality to generate one or more risk-sharing rules that are both fair and Pareto optimal, referred to here as \emph{actuarially fair Pareto optimal}. Now an \emph{actuarially fair Pareto optimal} risk-sharing rule can be characterized as follows.

\begin{corollary}[Characterization of AFPO risk-sharing rules]
\emph{
A risk-sharing rule $\left(h_{1}, \ldots, h_{n}\right)$ is actuarially fair Pareto optimal for any given total risk $S$ taking values in the domain $A$ if there exists a function $J: A \rightarrow \mathbb{R}_{+}$ and positive constants $\alpha_{1}, \ldots, \alpha_{n}$ such that
\begin{equation}
    \alpha_{i} v_{i}^{\prime}\left(h_{i}(s)\right)=J(s) \ \text{and} \ E\left[h_{i}(S)\right]=E[X_i] 
    \label{formulebis}
\end{equation}
for all $s \in A$ and for all $i \ \in \{1, \dots, n \}$.}
    \label{prob}
\end{corollary}
\noindent It should be noted that the distinction between Borch's Pareto optimality and Corollary \ref{prob} is that an additional fairness condition is incorporated as a constraint, which reduces the set of feasible schemes to a smaller class of fair risk-sharing schemes. Unlike the characterization of Pareto optimality by Borch, which does not require the distribution of $S$, here this distribution is important to establish the fairness condition.\\
It is straightforward that finding an AFPO risk-sharing rule is a root-finding procedure where one tries to find $\boldsymbol{\alpha}$ such that for all $i \ \in \ \{1, \dots, n\}$:
\begin{equation}
     E\left[I_{i}\left(J(S) / \alpha_{i}\right)\right]=E\left[X_i\right].
     \label{fairness}
\end{equation}
 From the expression \eqref{formule34} of $h_i$, one can see that multiplying $\boldsymbol{\alpha}$ by a constant, $J$ will also be multiplied by the same constant and thus $h_i$ remains unchanged.

\subsection{Examples}

Let us now have a look at some special cases in which Corollary \ref{prob} of the AFPO risk-sharing rule leads to known risk-sharing rules in the literature. The first one corresponds to the case where participants are homogeneous in risk aversion and losses.
\begin{example}[Uniform risk-sharing rule]
\emph{
    If a pool consists of $n$ participants who all have the same disutility function $v(s)$ and whose risk pool $\textbf{X} = (X_1, X_2, \dots, X_n)$ is exchangeable (i.e. joint probability distribution is invariant under permutation), then participants are indistinguishable. In this case, an appropriate risk-sharing rule is that each participant contributes an equal part of the aggregate loss ex-post and that the latter is thus uniformly distributed among all members of the pool. The unique risk-sharing rule $(h_1,...,h_n)$ is given by:}
\begin{equation}
    h_{i}(s)= \frac{s}{n}  \quad\forall \ i \ \in \{1, \dots, n \} \ \text{\emph{and for all}} \ s\ \in \ A.
    \label{formule62}
\end{equation}

In this framework, the unique AFPO risk-sharing rule of Example \ref{unif} is given by \eqref{formule62}. Indeed, let us assume that $\boldsymbol{\alpha}=(1,...,1)$. In this case, $h_i$ is given by \eqref{formule62}, and actuarial fairness is also verified. Now, it remains to show the uniqueness of the solution $\boldsymbol{\alpha}$. Thus, assume that the positive vector $(\alpha_1, . . . , \alpha_n)$ gives rise to a risk-sharing rule $h^{*}$ that is actuarially fair Pareto optimal. One denotes by $I(.)$ the inverse marginal disutility corresponding to the disutility function $v(s)$ and $Z=J(S;\boldsymbol{\alpha})$. The expression $h^{*}_{i}(s)=I_{i}\left(Z / \alpha_{i}\right)$ gives us the Pareto optimal risk-sharing rules. The actuarial fairness constraint leads to the following equality 
$E\left[I\left(Z / \alpha_{i}\right)\right]=C \ \forall \ i \ \in \ \{1, \ldots, n\}$, so that for all $i$ and $j$ one has
$$E\left[I\left(Z / \alpha_{i}\right)\right]=E\left[I\left(Z / \alpha_{j}\right)\right].$$
Since the inverse marginal disutility $I(.)$ is increasing, the function $ \boldsymbol{\alpha} \longmapsto E\left[I\left(Z / \alpha\right)\right]$ is decreasing. Thus, the equality above implies that $\alpha_i=\alpha_j$, and we find that all entries of $\boldsymbol{\alpha}$ must be equal.
\label{unif}
\end{example}

\begin{example}[Mean proportional risk-sharing rule\label{mpr}]
\emph{
  The mean proportional risk-sharing rule can be identified in a pool where all participants have equicautious CRRA (constant relative risk aversion) disutilities that is the same constant coefficient of risk aversion, so essentially all members have the same (power) disutility function. 
The unique AFPO solution $(h_1,...,h_n)$ is given in this case by:}

\begin{equation}
   h_i(s)=\frac{\mathbb{E}\left[X_i\right]}{\mathbb{E}\left[S\right]}\times s \quad\forall \ i \ \in \{1, \dots, n \} \ \text{and for all} \ s\ \in \ A. 
       \label{formule64}
\end{equation}
\emph{Participants thus agree to take a fixed percentage of the total loss $S$, in accordance with the expected values of the losses they bring to the pool compared to the total expected loss.}

To obtain this result, one can write the risk-sharing rules as a solution of a differential equation using Borch's condition for Pareto optimality (see \cite{eeckhoudt_economic_2005} (Section 10.3) for more details about the differential equation \eqref{diff1} below).
Fully differentiating the first-order condition (\ref{formule33}) yields $J^{\prime}(s)=\alpha_iv_i^{\prime \prime}(h_i(s))h_i^{\prime}(s).$
Eliminating $\alpha_i$ in this equation, using condition (\ref{formule33}), implies that        
$h_i^{\prime}(s)=T_i(h_i(s))\frac{J^{\prime}(s)}{J(s)}.$
Now, observe that fully differentiating the feasibility constraint
$\sum^{n}_{i=1}h_i(s)=s$
implies that
$\sum^{n}_{i=1}h_i^{\prime}(s)=1.$
Combining this with the above equation implies that
$$\frac{J^{\prime}(s)}{J(s)}\sum^{n}_{j=1}T_j(h_j(s))=1.$$
Finally, these last two equations together imply that
\begin{equation}
    h_i^{\prime}(s)=\frac{T_i(h_i(s))}{\sum^{n}_{j=1}T_j(h_j(s))}.
    \label{diff1}
\end{equation}
On the other hand, an equicautious CRRA disutility is characterized by a constant cautiousness (coefficient of relative risk aversion) denoted $\sigma$, such that for any $x$ the risk tolerance defined in \eqref{risktol} is given by $T_i(x) =\sigma x$ $\quad\forall \ i \ \in \{1, \dots, n \}$. The parameter $\sigma$ measures how quickly the coefficient of risk aversion decreases as risk goes down. This combines with the differential equation \eqref{diff1}, implies $h^{\prime}_i(s)=\frac{1}{s}h_i(s)$. The solution of such a differential equation is of the form $h_i(s)=k_i{s}$, where $\sum_{i=1}^{n}k_i=1$. It remains to determine $k_i$ by making use of the actuarial fairness constraint \eqref{formule32}. So, from $E\left[h_{i}(S)\right]=k_i{E(S)}=E(X_i)$ one deduces that $k_i=\frac{E(X_i)}{E(S)}.$
\label{exemple27}
\end{example}\bigskip

Examples \ref{unif} and \ref{exemple27} do not require a specific algorithm to compute the AFPO risk-sharing rules. However, for general disutility functions, Corollary \ref{prob} leads to a system of nonlinear equations that may not admit explicit solutions. We now propose a fast numerical method that solves risk-sharing rules of Corollary \ref{prob} for general disutility functions in  risk pooling.

\section{Fixed point approach for AFPO risk-sharing rules \label{sect3}}
As mentioned earlier, many nonlinear equations can be solved using fixed-point iteration. We will use this approach following the framework established by \cite{pazdera_composite_2017}. More precisely, we reformulate the problem of solving the system of equations consisting of the feasibility condition \eqref{self}, the actuarial fairness condition \eqref{formule32}, and the optimality condition \eqref{formule33} as a fixed-point problem. The variables that solve these equations are the disutility weights $\alpha_1, \dots, \alpha_n$, and the function $J$. There exists, at least, three main numerical methods implied for solving the system of equations which are:
\begin{enumerate}
    \item The Lagrange multiplier $J$ is fully determined by $\boldsymbol{\alpha}$ and $s$. Thus, the system \eqref{self}–\eqref{formule33} is equivalent to a system of $n$ nonlinear equations in $n$ variables $\alpha_1,...,\alpha_n$:
    \begin{equation}
         E\left[I_{i}\left(J(S; \alpha_1,...,\alpha_n) / \alpha_{i}\right)\right]=E[X_i] \quad\forall \ i \ \in \{1, \dots, n \}.
         \label{formule38}
    \end{equation}
Then, one can apply a nonlinear equation solver to find $\alpha_1,...,\alpha_n$;
    \item A second method consists in expressing disutility weights $\alpha_i$ in terms of the Lagrange multiplier $J$ by making use of the actuarial fairness conditions
\begin{equation}
    \sum_{i=1}^{n} I_{i}\left(J(s) / \alpha_{i}(J)\right)=s \ \text{for all} \ s\ \in \ A
    \label{formule39}
\end{equation}
for the function $J$;
    \item Finally, a third approach is to combine the two methods described above as follows: one uses the mapping of the first method to determine $J$ as a function of $\boldsymbol{\alpha}$ and the mapping of the second method to determine $\boldsymbol{\alpha}$ as a function of $J$. In this way, one can characterize the disutility weights that solve the equations system \eqref{formule38} as a fixed point of the composite of the previous mappings. Unlike the previous two approaches, this method is much less computationally demanding. 
\end{enumerate}
We focus on the third approach to propose a fast numerical method for computing AFPO solutions. Therefore, it is essential to study relevant properties of the composite iteration mapping to prove the existence, uniqueness, and convergence of the iterative algorithm. 

\subsection{Specification of relevant mappings}
Let $\boldsymbol{\alpha}$ and $\boldsymbol{\beta}$ in $\mathbb{R}^n$, and let us denote the following relations: $\boldsymbol{\alpha} \leq \boldsymbol{\beta}$ means that $\alpha_j\leq \beta_j$ for all $j=1, \dots, n$. Similarly, $\boldsymbol{\alpha} < \boldsymbol{\beta}$ means $\alpha_j < \beta_j$ for all $j=1, \dots, n$. . The non-negative cone $\{\boldsymbol{\alpha} \in  \mathbb{R}^n | \boldsymbol{\alpha} \geq 0 \}$ is denoted by $\mathbb{R}_{+}^n$, and $\mathbb{R}_{++}^n$ indicates the positive cone  $\{\boldsymbol{\alpha} \in  \mathbb{R}^n | \boldsymbol{\alpha} > 0 \}$.\\
Within the space of functions from $A$ to $\mathbb{R}_{+}$, equipped with the topology $\tau$ of pointwise convergence, one defines the cone of increasing functions
$$
\mathcal{K}=\left\{g \in \mathcal{L}(A,\mathbb{R}_{+}) \mid g(s)<g(v) \ \forall \ s, v \in A \text { s.t. } s<v \ \text {and} \ g(0)=0 \right\}.
$$

\subsubsection{Mapping from disutility weights $\alpha$ to $J$}
For $\boldsymbol{\alpha} \in \mathbb{R}_{++}^{n}$ and $z \in \mathbb{R}_{+}$, define
\begin{equation}
    F(z, \boldsymbol{\alpha})=\sum_{i=1}^{n} I_{i}\left(z / \alpha_{i}\right).
\end{equation}
This function is a continuous mapping from the product space $\mathbb{R}_{+} \times \mathbb{R}_{++}^{n}$ to $\mathbb{R}_{+}$. For a fixed $\boldsymbol{\alpha}$, the function $F(\cdot, \boldsymbol{\alpha})$ : $[0, +\infty) \rightarrow A$ is continuous and increasing, and satisfies
\begin{equation}
   \lim _{z \rightarrow \infty} F(z, \boldsymbol{\alpha})=\sum_{i=1}^{n} \max[X_i], \quad \lim _{z \rightarrow 0} F(z, \boldsymbol{\alpha})=0. 
\end{equation}
Thus, an inverse function exists for $F(. , \boldsymbol{\alpha})$ that is well-defined and is denoted by $J(. , \boldsymbol{\alpha})$.  $J(. , \boldsymbol{\alpha})$ is also continuous and increasing. For $\boldsymbol{\alpha} \in \mathbb{R}_{+}^{n}$, we can therefore define the function $\varphi_{1}(\boldsymbol{\alpha}) \in \mathcal{K}$ by

\begin{equation}
 \text{for} \ s \ \in A, \  \left(\varphi_{1}(\boldsymbol{\alpha})\right)(s)= \begin{cases}J(s, \boldsymbol{\alpha}) & \text { if } \boldsymbol{\alpha} > 0  \\ 0 & \text { otherwise.}\end{cases}
 \label{form53}
\end{equation}
In case $\boldsymbol{\alpha} > 0$, the function defining the mapping $\varphi_{1}$ can also be written in a form that is more implicit but probably more meaningful, i.e.
\begin{equation}
    \varphi_{1}: \boldsymbol{\alpha} \mapsto J, \quad \sum_{i=1}^{n} I_{i}\left(J(s) / \alpha_{i}\right)=s  \ \text{for all} \ s\ \in \ A.
\end{equation}

\subsubsection{Mapping from $J$ to disutility weights $\alpha$}
For any given function $J \in \mathcal{K}$, one defines for each $i=1, \dots, n$, the mapping $\alpha_i \mapsto E\left[I_{i}\left(J(S) / \alpha_{i}\right)\right]$. It is a decreasing function with
$$
\lim _{\alpha_i \rightarrow \infty} E\left[I_{i}\left(J(S) / \alpha_{i}\right)\right]=0, \quad \lim _{\alpha_i \rightarrow 0} E\left[I_{i}\left(J(S) / \alpha_{i}\right)\right]=\max[X_i].
$$
Since $E\left[X_i\right] < \max[X_i]$, the equation
\begin{equation}
  E\left[I_{i}\left(J(S) / \alpha_{i}\right)\right]=E\left[X_i\right] 
  \label{formule55}
\end{equation}
 has a unique solution $\alpha_{i} > 0$. The mapping from $J$ to the disutility weights $\alpha$ can be extended to a mapping defined on all elements of $\mathcal{K}$ by
\begin{equation}
    \left(\varphi_{2}(J)\right)_{i}= \begin{cases}\alpha_{i} \ \text{satisfying} \ \eqref{formule55} & \text { if } J \neq 0 \\ 0 & \text { if } J=0\end{cases}
    \label{nolab}
\end{equation}
for $i=1, \ldots, n.$

\subsubsection{The complete iteration mapping}
With the mappings $\varphi_1: \mathbb{R}_{+}^{\mathrm{n}} \rightarrow \mathcal{K}$ and $\varphi_2: \mathcal{K} \rightarrow \mathbb{R}_{+}^{\mathrm{n}}$ in hand, one can define a mapping $\varphi$ from $\mathbb{R}_{+}^{\mathrm{n}}$ into itself by

\begin{equation}
    \varphi(\boldsymbol{\alpha})=\varphi_{2}\left(\varphi_{1}(\boldsymbol{\alpha})\right).
    \label{form519}
\end{equation}

As mentioned above, there is a \emph{one-to-one correspondence} between an AFPO solution and a fixed point of $\varphi$. The main contribution of this paper is to prove the existence and uniqueness of the fixed point of $\varphi$ and to provide a fast numerical algorithm to compute it.

To fix the idea of the algorithm, we give an analytical illustration of the fixed point approach in the following example. The proof of this example is given in Appendix \ref{A15}.

\begin{example}
\emph{
    We assume a pool of two participants ($n=2$) each carrying a risk $X_i$ with $i=1,2$. The disutility functions of the two agents are of the CRRA type, i.e. $v_i(s)=\frac{s^{1+\sigma_i}}{1+\sigma_i}$ with $\sigma_2=2 \sigma_1$. The inverse of the marginal disutility of participant $i$ is therefore in the form $I_i(z)=z^{\frac{1}{\sigma_i}}.$}\newline
    \begin{enumerate}
        \item \emph{ \textbf{Specification of the mapping $\varphi_{1}$} }
       $$
            \left(\varphi_{1}(\boldsymbol{\alpha})\right)(s)= \frac{\left(\alpha_1\right)^2}{4^{\sigma_1}}\left(\sqrt{\left(\alpha_2\right)^{-1 / \sigma_1}+4 s\left(\alpha_1\right)^{-1 / \sigma_1}}-\left(\alpha_2\right)^{-1 / 2 \sigma_1}\right)^{2 \sigma_1} \ \text{with} \ \boldsymbol{\alpha}=\left(\alpha_1, \alpha_2 \right);
        $$
        
        \item \emph{ \textbf{Specification of the mapping $\varphi_{2}$} }

        $$
                    \left(\varphi_{2}(J)\right)=\left(\left(\frac{\mathbb{E}[\left(J\left(S\right)\right)^{\frac{1}{\sigma_1}}]}{\mathbb{E}[X_1]}\right)^{\sigma_1}, \left(\frac{\mathbb{E}[\left(J\left(S\right)\right)^{\frac{1}{\sigma_2}}]}{\mathbb{E}[X_2]}\right)^{\sigma_2}\right);
        $$

         \item \emph{ \textbf{Fixed point} $\boldsymbol{\tilde{\alpha}}= \left(\tilde{\alpha_1}, \tilde{\alpha_2}\right)$  \textbf{such that} $\varphi_2\left(\varphi_1(\boldsymbol{\tilde{\alpha}})\right)=\boldsymbol{\tilde{\alpha}}$ \textbf{is given by} }

        $$
                \tilde{\alpha_1}=\frac{a^{\sigma_1}}{1+a^{\sigma_1}}, \quad \tilde{\alpha_2}= \frac{1}{1+a^{\sigma_1}}
                \label{alpha1}
        $$
            
            \emph{where $a$ is the unique solution of the equation}
            
        $$
             \mathbb{E}\left[\sqrt{a S + \left(\frac{a}{2}\right)^{2} }-\frac{a}{2}\right]= \mathbb{E}\left[X_2\right];
             \label{fixpoint}
        $$

    \item \emph{ \textbf{The corresponding AFPO risk-sharing rule }is}

    \begin{equation}
    h_2(s)= I_2\left(\frac{J\left(s\right)}{\tilde{\alpha_2}}\right)= \sqrt{a s + \left(\frac{a}{2}\right)^{2} }-\frac{a}{2};
    \label{h2_nonl}
\end{equation}

\begin{equation}
    h_1(s)= s - h_2(s)= s- \sqrt{a s + \left(\frac{a}{2}\right)^{2} }+\frac{a}{2}.
    \label{h1_nonl}
\end{equation}
\label{toym}
 \end{enumerate}
\end{example}

\subsection{Existence and uniqueness of the fixed point \label{part1}}
The techniques for establishing existence and uniqueness are inspired by \cite{pazdera_composite_2017}. However, due to differences in assumptions, we rely on different arguments than those used by \cite{pazdera_composite_2017} to prove the desired results.

Let's start with some notations. When $\boldsymbol{\alpha}$ is a given vector in $\mathbb{R}^n$ and $Q = \{i_1, \dots, i_k\}$ is a nonempty subset of the index set $\{1, \dots, n\}$, we write $\boldsymbol{\alpha}_Q := (\alpha_{i_1}, \dots, \alpha_{i_k})$.
A fixed point being an eigenvector associated with an eigenvalue equal to $1$, we rely on Oshime's theorem \cite{oshime_extension_1983} below to study the existence and uniqueness of the fixed point of $\varphi$.

\begin{definition}
\emph{
The mapping $\varphi$ of $\mathbb{R}_{+}^{\mathrm{n}}$ is called non-sectional if it satisfies the following:
for any given decomposition of the index set $\{1, \ldots, n\}$ into two complementary nonempty subsets $Q$ and $R$ there exists $q \in Q$ for which following 1 and 2 holds at the same time:
\begin{enumerate}
    \item $\left(\varphi\left(\boldsymbol{w}\right)\right)_q > \left(\varphi\left(\boldsymbol{y}\right)\right)_q$ for all $\boldsymbol{w}, \boldsymbol{y} \in \mathbb{R}_{+}^{\mathrm{n}}$ such that $\boldsymbol{w}_R > \boldsymbol{y}_R$ and $\boldsymbol{w}_Q=\boldsymbol{y}_Q > 0$;
    \item $\left(\varphi\left(\boldsymbol{w}^{k}\right)\right)_q \rightarrow \infty$ for all sequences $\left(\boldsymbol{w}^{k}\right)_{k=1,2, \ldots} \in \mathbb{R}_{+}^{\mathrm{n}}$ such that $\boldsymbol{w}_{R}^{k} \rightarrow \infty$ while $\boldsymbol{w}_{Q}^{k}$ is fixed and positive.
\end{enumerate}
}
\label{def4}
\end{definition}

\begin{theorem}
\emph{
(\cite{oshime_extension_1983}, Thm. 8, Remark 2). If the mapping $\varphi$ from $\mathbb{R}_{+}^{\mathrm{n}}$ into itself is continuous, monotone, homogeneous, and \emph{non-sectional},
then the mapping $\varphi$ has a positive eigenvector, which is unique up to a scalar multiplication, with a positive associated eigenvalue. In other words, there exists a constant $\lambda^* > 0$ and a vector $s^* \in \mathbb{R}_{++}^{\mathrm{n}}$ such that $\varphi(s^*)=\lambda^*{s^*}$. Further if $\lambda > 0$ and $s \in \mathbb{R}_{++}^{\mathrm{n}}$ are such that
$\varphi(s)=\lambda{s}$, then $s$ is a scalar multiple of $s^*$.}
\label{osh}
\end{theorem}
\noindent It is shown in (\cite{oshime_extension_1983}, Thm. 3) that the eigenvalue associated with the positive eigenvector of the above theorem is in fact the maximum eigenvalue of the mapping $\varphi$. Thus, this positive eigenvector can be obtained by iteration.
In the following, one shows that $\varphi$ satisfies the properties in Oshime's Theorem \ref{osh}. To do so, the properties of the mappings $\varphi_{1}$ and $\varphi_{2}$ and $\varphi=\varphi_{2}\circ\varphi_{1}$ are studied separately.\bigskip

\noindent All the proofs of Section \ref{part1} are gathered in Appendix \ref{A2}.

\subsubsection{Properties of $\varphi_{1}$}
In this part, one shows that the mapping $\varphi_{1}$ verifies some properties that are required to apply Oshime's Theorem \ref{osh}.

\begin{lemma}
\emph{The mapping $\varphi_{1}$ is homogeneous, monotone, and continuous.}
   \label{lemma36}
\end{lemma}
\noindent The next lemma states a property of the mapping $\varphi_{1}$ that relates to non-sectionality.

\begin{lemma}
\emph{
Let $\left(\boldsymbol{\alpha}^{k}\right)_{k=1,2, \ldots}$ be a sequence in $\mathbb{R}_{+}^{\mathrm{n}}$ that has the following property: there exist complementary nonempty index sets $Q$ and $R$ in $\{1, \ldots, n\}$ and a vector $\boldsymbol{\alpha}_{Q} \in \mathbb{R}_{++}^{|Q|}$ such that $\boldsymbol{\alpha}_{R}^{k} \rightarrow \infty$ as $k \rightarrow \infty$, while $\boldsymbol{\alpha}_{Q}^{k}=\boldsymbol{\alpha}_{Q}$ for all $k$. Then $\left(\varphi_{1}\left(\boldsymbol{\alpha}^{k}\right)\right)(s) \rightarrow \infty$ as $k \rightarrow \infty$ for all $s \in A$.}
\label{lemma54}
\end{lemma}

\subsubsection{Properties of $\varphi_{2}$}
In this part, one shows that the mapping $\varphi_{2}$ verifies some properties that are required to apply Oshime's Theorem \ref{osh}.

\begin{lemma}
\emph{
    The mapping $\varphi_{2}$ is homogeneous, monotone, and sequentially continuous.}
    \label{lemma38}
\end{lemma}
\noindent The final lemma establishes a property that will be used in a non-sectionality argument.

\begin{lemma}
\emph{
Let $\left(J_{k}\right)_{k=1,2, \ldots}$ be a sequence in $\mathcal{K}$ such that $J_{k}(s) \rightarrow \infty$ for all $s \in A$ as $k \rightarrow \infty$. Then $\left(\varphi_{2}\left(J_{k}\right)\right)_{i} \rightarrow \infty$ for all $i=1, \ldots, n$}
\label{lemma57}.
\end{lemma}

\subsubsection{Properties of $\varphi$}
In this part, the results of the properties of $\varphi_{1}$ and $\varphi_{2}$ are used to establish the existence and uniqueness of the fixed point of $\varphi$.

\begin{proposition}
\emph{
The mapping $\varphi$ can only have 1 as an eigenvalue corresponding to a positive eigenvector. In other words, if $\alpha \in \mathbb{R}_{++}^{n}$ is such that $\varphi(\alpha)=\lambda \alpha$, then $\lambda=1$.}
\label{prop1}
\end{proposition}

\begin{theorem}
\emph{
The mapping $\varphi$ has a unique continuous extension to a mapping from the nonnegative cone to itself. This extension is homogeneous, strictly monotone, and non-sectional. On the positive cone, the mapping $\varphi$ is strongly monotone.}
\label{theo1}
\end{theorem}

One can conclude using Oshime's Theorem \ref{osh} above.

\begin{corollary}
\emph{
The system of equations characterizing the AFPO risk-sharing rules given in Corollary \ref{prob} has a unique solution. The unique risk-sharing rule that is actuarially fair Pareto optimal is given by
$h_{i}(s)=I_{i}\left(J(s) / \alpha_{i}\right)$
for $i=1, \ldots, n$ and $s \in A$, where $I_{i}$ is the inverse marginal disutility function of participant $i$, $\boldsymbol{\alpha}=\left(\alpha_{1}, \ldots, \alpha_{n}\right)$ is a positive eigenvector (unique up to multiplication by a positive scalar) of the mapping $\varphi$ and $J$ is given by $J=\varphi_{1}(\boldsymbol{\alpha})$.}
\end{corollary}

\subsection{Convergence of the fixed point iterative algorithm \label{part2}}
As in \cite{pazdera_composite_2017}, we rely on Nadler's theorem to establish the convergence of the fixed-point algorithm.

The fixed point we are looking for is in fact associated with the maximum eigenvalue of $\varphi$ thanks to Oshime (\cite{oshime_extension_1983}, Thm. 3). Thus, this fixed point can be obtained by iteration.
In the following, the convergence of the fixed point iterative algorithm is studied. For this purpose, we begin by defining a normalized mapping $\psi$ from the open unit simplex $\left\{\boldsymbol{s}=\left(s_{1}, \ldots, s_{n}\right) \in \mathbb{R}_{++}^{n} \mid \sum_{i=1}^{n} s_{i}=1\right\}$ into itself by
\begin{equation}
    \psi(\boldsymbol{s})=\frac{\varphi(\boldsymbol{s})}{\|\varphi(\boldsymbol{s})\|_{1}}
    \label{psi}
\end{equation}
where $\|\boldsymbol{v}\|_{1}=\sum_{i=1}^{n}\left|v_{i}\right|$ is the 1-norm of $\boldsymbol{v} \in \mathbb{R}^{n}$. The fixed point of the mapping $\varphi$ corresponds also to the fixed point of the mapping $\psi$. A standard metric used for this kind of study is the Hilbert metric, which is defined as follows.

\begin{definition}
\emph{
The Hilbert metric assigns to a pair $(\boldsymbol{w}, \boldsymbol{y})$ with $\boldsymbol{w}, \boldsymbol{y} \in \mathbb{R}_{++}^{\mathrm{n}}$ the distance $d(\boldsymbol{w}, \boldsymbol{y})$ given by
$$d(\boldsymbol{w}, \boldsymbol{y})=\log\frac{\max_i(w_i/y_i)}{\min_i(w_i/y_i)}.$$
}
\end{definition}
\noindent Points on the same ray are equivalent with respect to the Hilbert metric, since
\begin{equation}
    d(a\boldsymbol{w}, b\boldsymbol{y})= d(\boldsymbol{w}, \boldsymbol{y}) \ \text{for all} \ a, \ b \ > 0.
    \label{prop41}
\end{equation}

\noindent All the proofs of Section \ref{part2} are gathered in Appendix \ref{A1}.

\begin{lemma}
\emph{
If $ \varphi : \mathbb{R}_{++}^{\mathrm{n}} \rightarrow \mathbb{R}_{++}^{\mathrm{n}}$ is homogeneous and strongly monotone, then $\varphi$ is contractive with respect to the Hilbert metric i.e., $d(\varphi(\boldsymbol{w}), \varphi(\boldsymbol{y})) < d(\boldsymbol{w}, \boldsymbol{y})$ for all $\boldsymbol{w}, \boldsymbol{x}$ such that $d(\boldsymbol{w}, \boldsymbol{y}) > 0$.}
\label{lemma44}
\end{lemma}
\noindent One can see that if $\varphi$ is contractive then $\psi$ is also contractive because of property \eqref{prop41}.

\begin{theorem}
\emph{
(\cite{nadler_jr_note_1972}, Thm. 1).
If $(\mathcal{X}, d)$ is a locally compact and connected metric space, and $f : \mathcal{X}  \rightarrow \mathcal{X}$ is a contractive mapping with fixed point $x^* \in \mathcal{X}$, then for every $x \in \mathcal{X}$ the sequence of iterates $(f^{(k)}(x))_{k=1,2,...}$ converges to the point $x^*$.}
\label{nadler}
\end{theorem}

\noindent Using the fact that the open unit simplex in finite dimensions is a locally compact and connected metric space, we can therefore apply Theorem \ref{nadler} to conclude the global convergence of the composite iteration algorithm.

\begin{corollary}
\emph{
The mapping $\psi$ has the following property: for every $\boldsymbol{\alpha}^{0}$ in the open unit simplex $\left\{\boldsymbol{\alpha} \in \mathbb{R}_{++}^{n} \mid \sum_{i=1}^{n} \alpha_{i}=1\right\}$, the sequence of vectors $\left(\boldsymbol{\alpha}^{0}, \boldsymbol{\alpha}^{1}, \ldots\right)$ defined iteratively by
\begin{equation}
  \boldsymbol{\alpha}^{i+1}=\psi\left(\boldsymbol{\alpha}^{i}\right)
  \label{formule58}
\end{equation}
converges to the unique eigenvector in the open unit simplex of the mapping $\varphi$.}
\end{corollary}

\noindent Algorithm \ref{algo} gives a description of the numerical algorithm for computing AFPO risk-sharing rule.
\begin{algorithm}
    \SetKwFunction{isOddNumber}{isOddNumber}
    \SetKwInOut{KwIn}{Input}
    \SetKwInOut{KwOut}{Output}

    \KwIn{Distribution of $X_i$ and participants’ preferences specified by $v_i$, $i=1, \cdots, n$.}
    \KwOut{$h_i(s) \ \forall \ s \ \in \ A$ for each $i=1, \cdots, n$.}
    Compute $I_i(s)=\left(v_i^{\prime}\right)^{-1}(s)$ $\forall$ $s$ $\in$ $A$ for each $i=1, \cdots, n$;
    
    Compute the distribution of $S=\sum_{i=1}^{n}X_i$;
    
    Start with some initial $\boldsymbol{\alpha}^{(0)} \in \mathbb{R}_{++}^{n}$;
    
    For any given $\boldsymbol{\alpha}^{(m)}$ with $m \in \mathbb{N}$, determine $J^{(m)}(s, \boldsymbol{\alpha}^{(m)})$ by solving for any $s$ $\in$ $A$ the following equation:
    $$
                        \sum_{i=1}^{n} I_{i}\left(\frac{J^{(m)}(s)}{\alpha_{i}^{(m)}}\right)=s \text {; }
                        $$
    
    Update the $\boldsymbol{\alpha}$ from $\boldsymbol{\alpha}^{(m)}$ to $\boldsymbol{\alpha}^{(m+1)}$ by solving for each $i=1, \cdots, n$ the following equation:
                            $$
                            \mathbb{E}\left[I_{i}\left(\frac{J^{(m)}\left(S, \boldsymbol{\alpha}^{(m)}\right)}{\alpha_{i}^{(m+1)}}\right)\right]=\mathbb{E}\left[X_{i}\right];
                            $$
                            
    Compute $J^{(m+1)}(s, \boldsymbol{\alpha}^{(m+1)})$ $\forall \ s \ \in \ A$ using step 3;
    
    Compute the risk-sharing rules $\forall \ s \ \in \ A$ for each $i=1, \cdots, n$ by
    $$h_i^{(m+1)}(s)=I_{i}\left(\frac{J^{(m+1)}(s, \boldsymbol{\alpha}^{(m+1)})}{\alpha_{i}^{(m+1)}}\right);$$
    
    If for some pre-specified error tolerance $\epsilon$ 
    $$
       \|\boldsymbol{\alpha}^{(m+1)} - \boldsymbol{\alpha}^{(m)}\|_{2}  < \epsilon \ \ \ \tcp*[f]{Euclidean distance error} ;   
    $$
    conclude that $\left(h_1^{(m+1)}, \ldots, h_n^{(m+1)}\right)$ is the AFPO risk-sharing rule. Otherwise, go to step 3 with $\boldsymbol{\alpha}^{(m+1)}$;
    
    \KwRet{$\left(h_1, \ldots, h_n\right).$}
    \caption{Computing AFPO risk-sharing rules}
    \label{algo}
\end{algorithm}
\begin{remark}
In the case where the risk tolerance defined in \eqref{risktol} is of the form $T_i(s) = \sigma{s}+\tau_i$ for $i \ \in \ \{1, \dots, n\}$, Algorithm \ref{algo} converges in one step. The proof is given in Appendix \ref{A14} and follows the same arguments as the ones of Proposition 6.1 in \cite{pazdera_composite_2017}.
\label{prop61} 
\end{remark}

\section{Numerical illustrations of the fixed point algorithm\label{sect4}}
In this section, some numerical illustrations using the fixed point algorithm to obtain AFPO risk-sharing rules are presented. We start with Example \ref{mpr} of the mean proportional risk-sharing rule above, where the solution is known. Next, the sensitivity analysis of the risk-sharing rules is performed based on the risk aversion of the participants. Finally, large portfolios of risks are considered to show that our numerical method scales well to situations with many participants. 

To implement the different applications below, we present the global computation framework that has been adopted in this section.
As a reminder, the inputs of our algorithm are:
\begin{enumerate}
    \item participants’ risks aversion, specified by $v_i$;
    \item the joint distribution of individual risks which they bring into the pool.
\end{enumerate}
The crucial step of this algorithm is to compute the distribution of $S=\sum^{n}_{i=1} X_i$ using the joint distribution $\left(X_1, \dots, X_n\right)$. For all the examples below, one uses the fast Fourier transform (FFT) algorithm to determine the distribution of the random variable $S$. FFT provides an efficient method to extract the values of ordinary generating functions. It has been demonstrated in \cite{embrechts_panjer_2009} that computing the probability mass function of a compound sum is more efficient with the FFT than using the Panjer recursion method or by direct convolution. When using the FFT algorithm for a discrete random variable, one important factor to consider is selecting a truncation point that is large enough such that $f_S(k_{max})=0$ for $k_{max}$ the maximal value of $S$. A large value of $k_{max}$ may be required if $S$ represents a large portfolio or if the individual risks have heavy tails. In order to apply the generating function approach (see \cite{blier-wong_generating_2022} for more details of this approach) with the FFT algorithm (or other efficient convolution algorithms) to continuous random variables, it is necessary to discretize their continuous cumulative distribution functions with a step size $\delta \in \mathbb{R}_+$. For a brief overview of the upper, lower, and mean preserving discretization methods and their applications using the FFT algorithm, we refer to Section 2 of \cite{embrechts_panjer_2009}.

To find the fixed point, we begin with an initial value $\alpha^{(0)}$ and iterate the mapping $\varphi$ until convergence. In all the examples below, a uniform starting point has been chosen, i.e. $\alpha^{(0)}= (\frac{1}{n}, \dots, \frac{1}{n})$. In the $(m+1)-th$ iteration, the Euclidean distance below is evaluated:

        \begin{equation}
            \|\boldsymbol{\alpha}^{(m+1)} - \boldsymbol{\alpha}^{(m)}\|_{2}.
            \label{myfeasib2}
        \end{equation}
This quantity allows specifying a tolerance $\epsilon$ (fixed to $10^{-14}$ in our numerical illustrations), beyond which one can stop the iteration process without finding the exact fixed point.

\subsection{Equicautious CRRA pool \label{crra}}
It is known in Example \ref{mpr} that an equicautious CRRA pool leads theoretically to a well-known risk-sharing rule: the mean proportional risk-sharing rule. Here, we apply the fixed point algorithm in such a pool. \\
Consider an equicautious CRRA pool, in this case, participants have the same disutility function given by: 
\begin{equation}
  v(s)=\frac{s^{1+\sigma}}{1+\sigma}, 
\end{equation}
where $\sigma=2$ (in this example) is the cautiousness or the coefficient of relative risk aversion of participants. In this case, the fixed point algorithm must always yield risk-sharing rules identical to the expression

\begin{equation}
   h_i(s)=\frac{\mathbb{E}\left[X_i\right]}{\mathbb{E}\left[S\right]}\times s \quad\forall \ i \ \in \{1, \dots, n \} \ \text{and for all} \ s\ \in \ A.
\end{equation}
To fix the idea, let us assume that the loss distribution $X_i$ follows a compound Poisson distribution with parameter $\lambda_i$ and a Negative Binomial severity (NBinom$(r_i, q_i)$). The parameters used for this example are presented in Table \ref{param4} below.
\begin{table}[h!]
            \centering
           
            \begin{tabular}{|c|c|c|c|c|c|c|c|c|}
                \hline
                Participant $i$ & 1 & 2 & 3 & 4\\
                \hline
                
                $\lambda_i$ & 0.13 & 1.97 & 0.26 & 0.01 \\

                $r_i$ & 1 & 4 & 4 & 4\\

                $q_i$ & 0.41 & 0.47 & 0.41 & 0.45\\
                $\mathbb{E}[X_i]$ & 0.13 & 1.97 & 0.26 & 0.01\\
                \hline
                $\boldsymbol{\alpha}$ & $1.25\times10^{-2}$ & $5.32\times10^{-5}$ & $3.10\times10^{-3}$ & $9.84\times10^{-1}$\\ 
                \hline                
            \end{tabular}
            \caption{Marginal parameters of the equicautious CRRA pool.}
            \label{param4}
\end{table}
The probability mass function of $S$ is represented in Figure \ref{density_mpr} below.

\begin{figure}[h!]
        \centering
                \includegraphics[width=.5\linewidth]{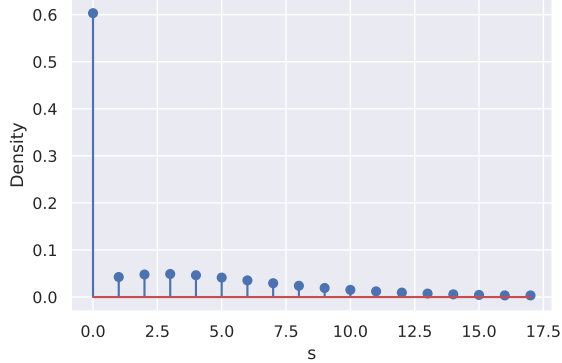}
                \caption{Probability mass function of the aggregated random variable $S$.}
                \label{density_mpr}
\end{figure}
\begin{figure}[h!]
            \centering
                    \includegraphics[width=.6\linewidth]{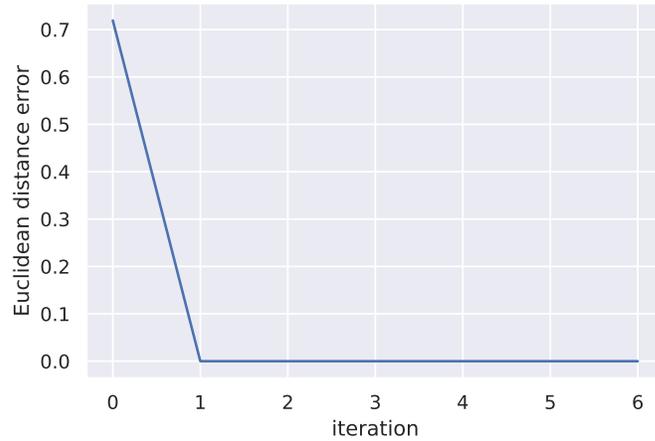}
                    \caption{Euclidean distance as a function of the number of iterations.}
                    \label{conv_hpr}
    \end{figure}
\begin{figure}[h!]
        \centering
            \begin{minipage}{0.48\linewidth}
                \centering
                \includegraphics[width=.8\linewidth]{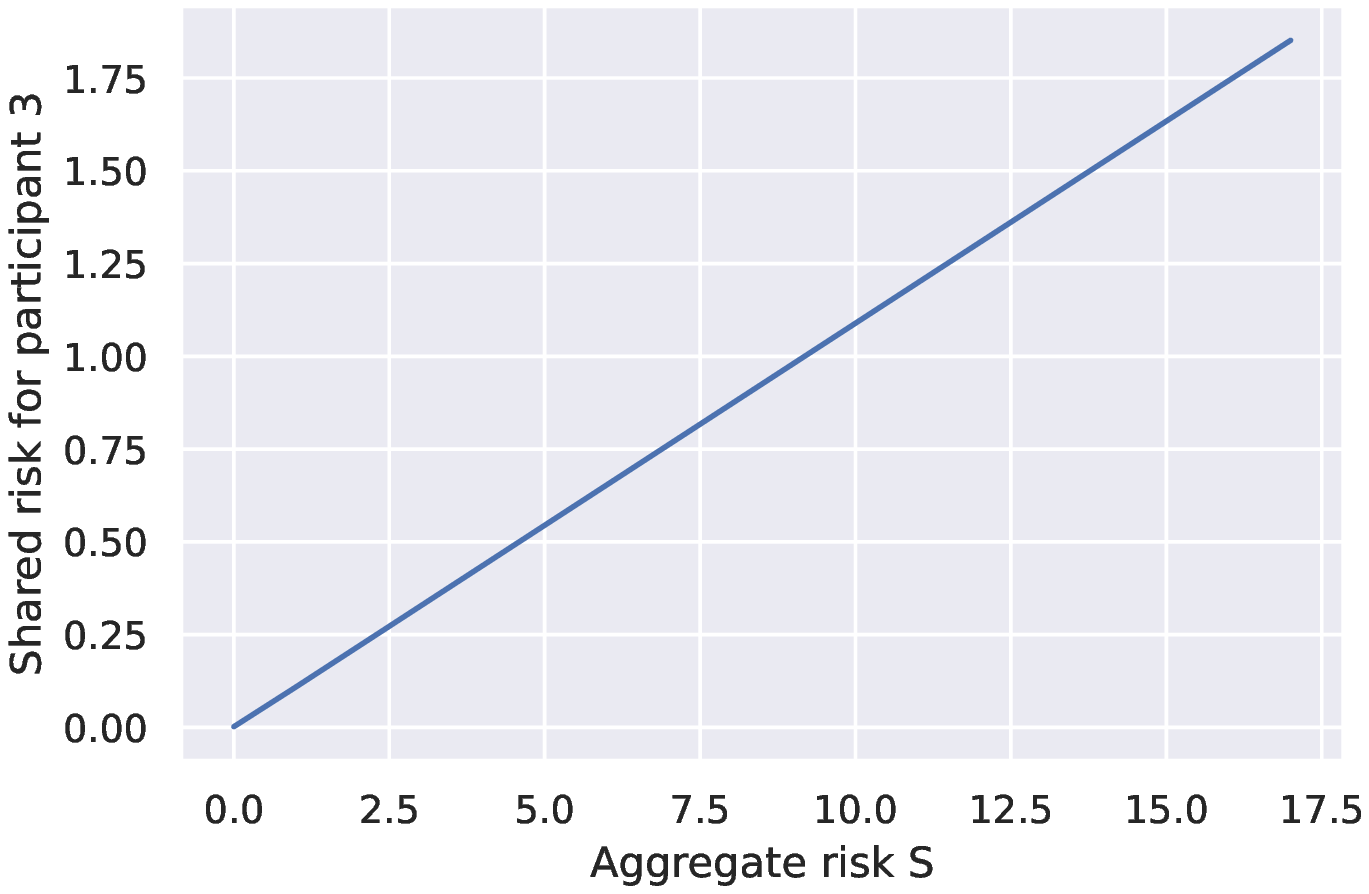}
            \end{minipage}%
            \hspace{4pt}
            \begin{minipage}{0.48\linewidth}
                \centering
                \includegraphics[width=.8\linewidth]{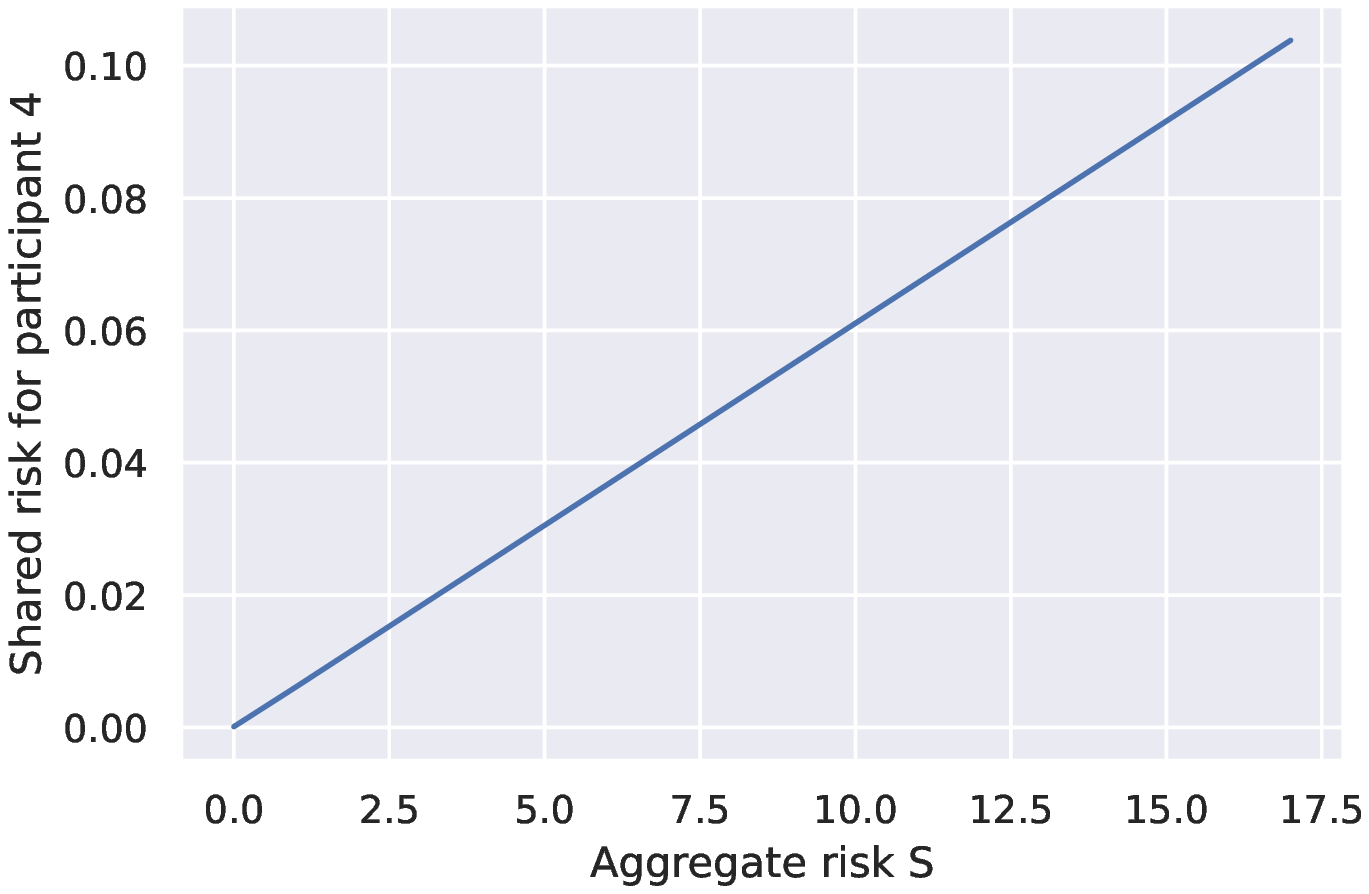}
                
            \end{minipage}
             \caption{Risk-sharing rules of the four participants of the equicautious CRRA pool.}
             \label{hpr4}
        \end{figure}

\noindent The iterative algorithm, which starts at the point $\alpha^0=(\frac{1}{4},\frac{1}{4}, \frac{1}{4}, \frac{1}{4})$, produces a solution in one iteration that satisfies the Euclidean distance constraint \eqref{myfeasib2} (see Figure \ref{conv_hpr}). This result confirms Remark \ref{prop61} which stated that in \emph{equicautious CRRA} pool, Algorithm \ref{algo} converges in one step. The fixed point for these four participants is given in Table \ref{param4}. We observe that participant 4, who has the lowest expected loss, is assigned a much larger disutility weight of $9.84\times10^{-1}$ compared to the other participants. On the other hand, participant 2 with the highest expected loss is assigned the smallest disutility weight of $5.32\times10^{-5}$. Figure \ref{hpr4} represents the evolution of the risk-sharing rules $h_i$ ($i=1, \dots, 4$) as a function of the aggregated risk $S$. It can be noted that all the functions $h_i$ are affine. Furthermore, the approximation error is almost zero when comparing the values of these risk-sharing rules to those obtained with the closed-form formula. This result validates our algorithm for this example.

       \begin{figure}[h!]
        \centering
            \begin{minipage}{0.48\linewidth}
                \centering
                \includegraphics[width=.8\linewidth]{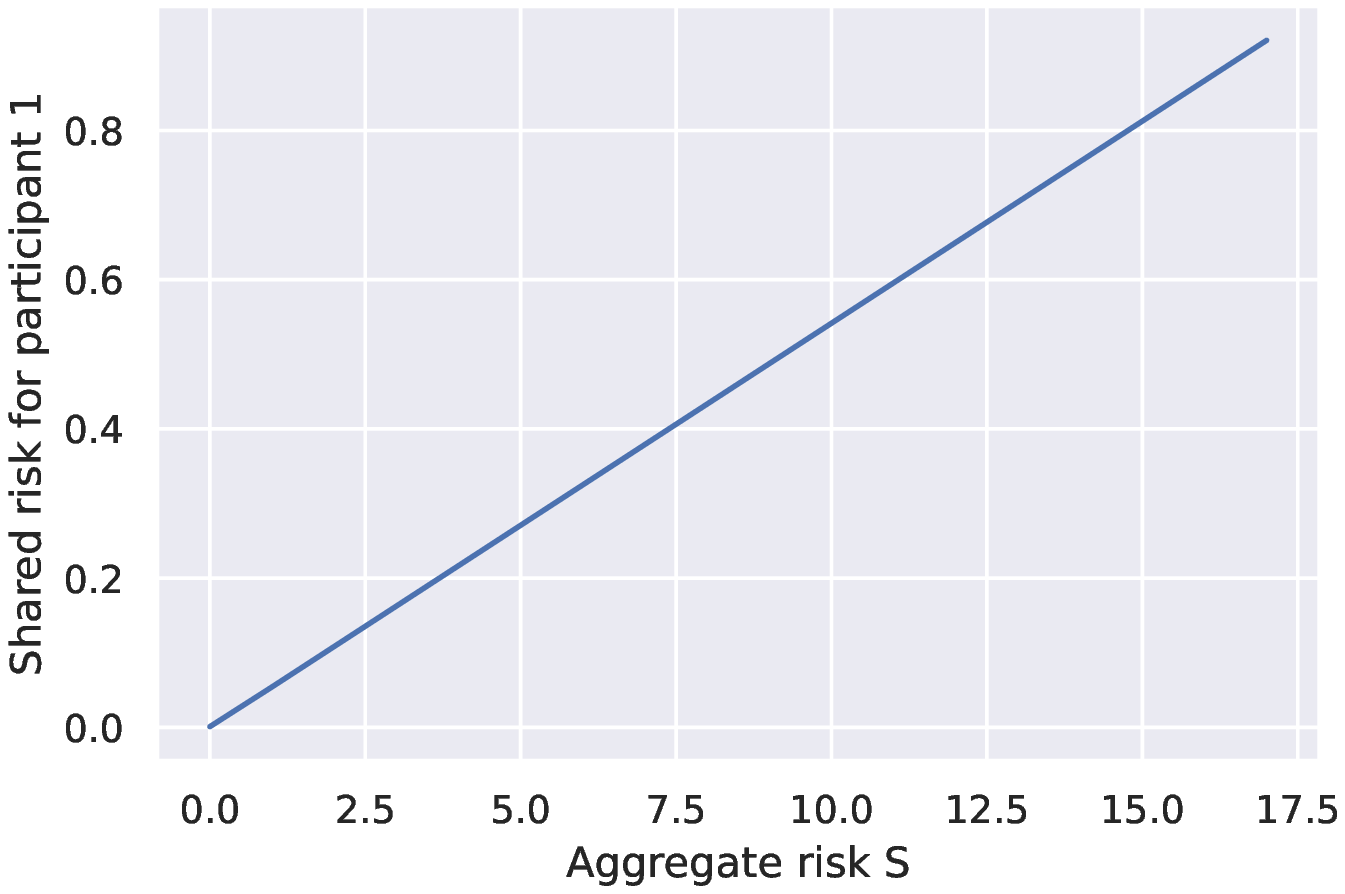}
            \end{minipage}%
            \hspace{4pt}
            \begin{minipage}{0.48\linewidth}
                \centering
                \includegraphics[width=.8\linewidth]{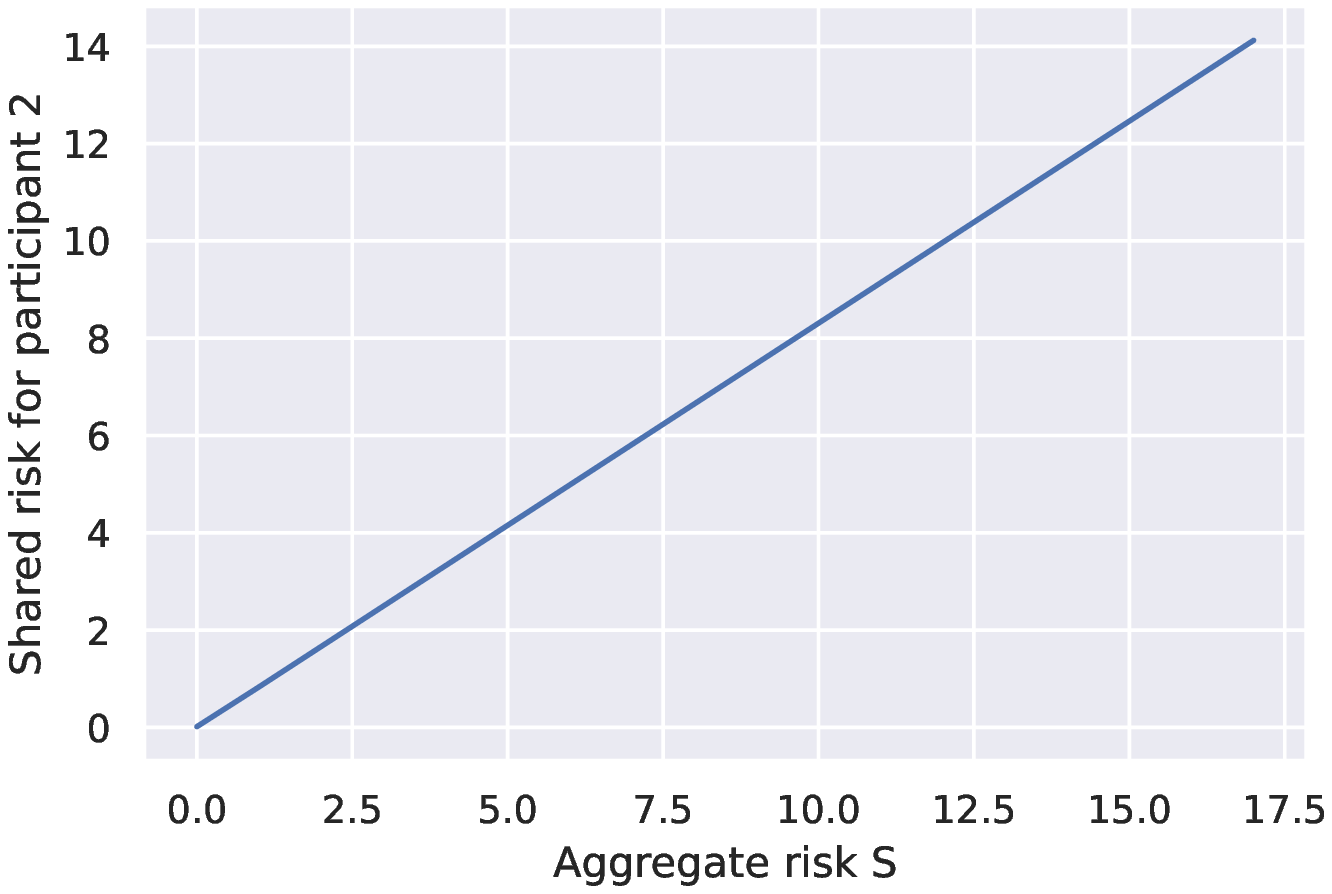}
            \end{minipage}
        \end{figure}

\subsection{Sensitivity of the risk-sharing rules to risk aversion parameters \label{mysen}}
In this section, we discuss the effects of participants' risk aversion on the risk-sharing rule. To do this, one considers the example in Section \ref{crra} above (equicautious CRRA pool) where the coefficient of relative risk aversion ($\sigma$) and the distribution of the losses are changed. Indeed, in the example of the equicautious CRRA pool, all participants have the same $\sigma$ fixed to $2$ which leads to affine risk-sharing rules for all of them. Table \ref{param4bis} shows each participant's values of $\sigma$. Also, to capture only the impact of $\sigma_i$, identical loss distributions are used for all participants. Thus we set $\lambda_i=0.26$, $r_i=4$ and $q_i=0.47$ for all $i=1, \dots, n.$  

\begin{figure}[h!]
        \centering
            \begin{minipage}{0.48\linewidth}
                \centering
                \includegraphics[width=.8\linewidth]{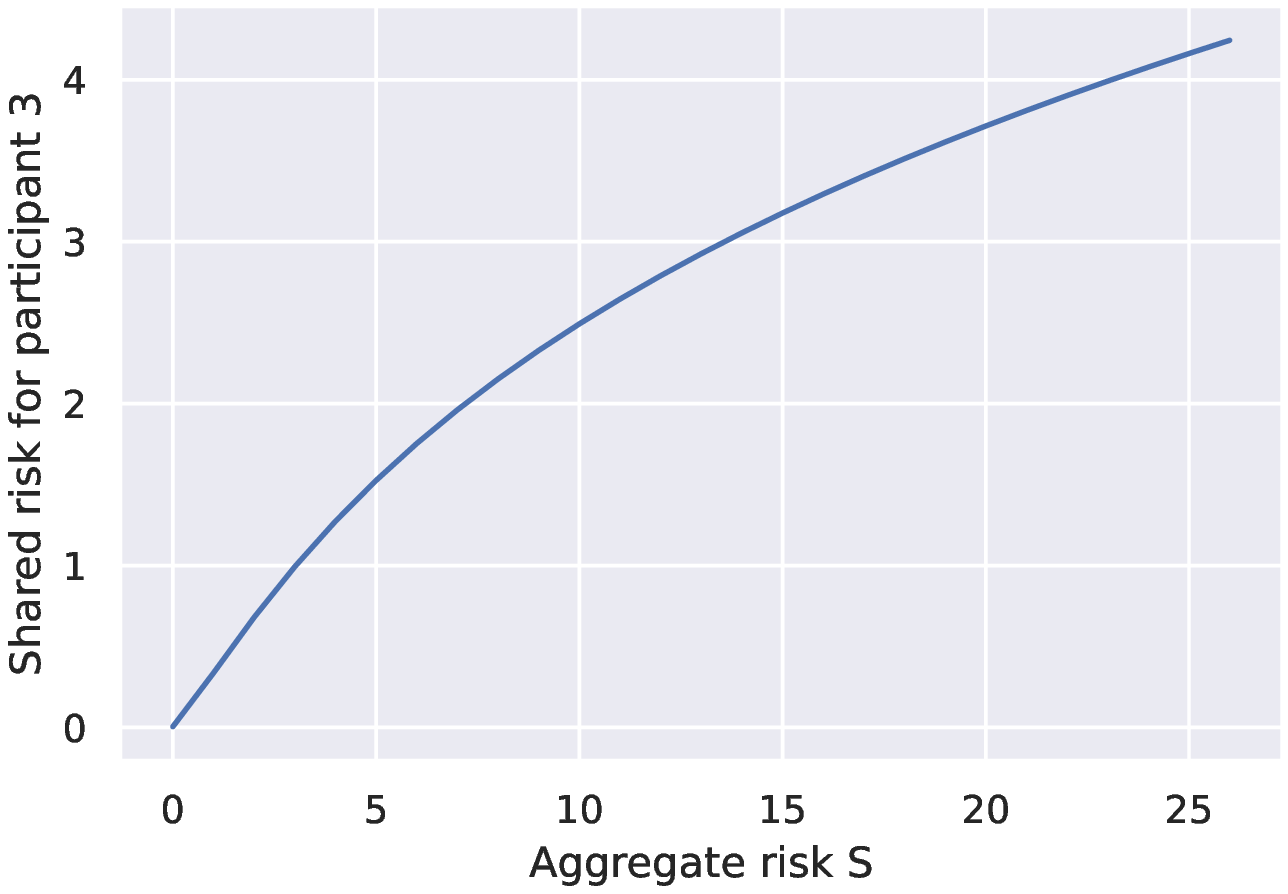}
            \end{minipage}%
            \hspace{4pt}
            \begin{minipage}{0.48\linewidth}
                \centering
                \includegraphics[width=.8\linewidth]{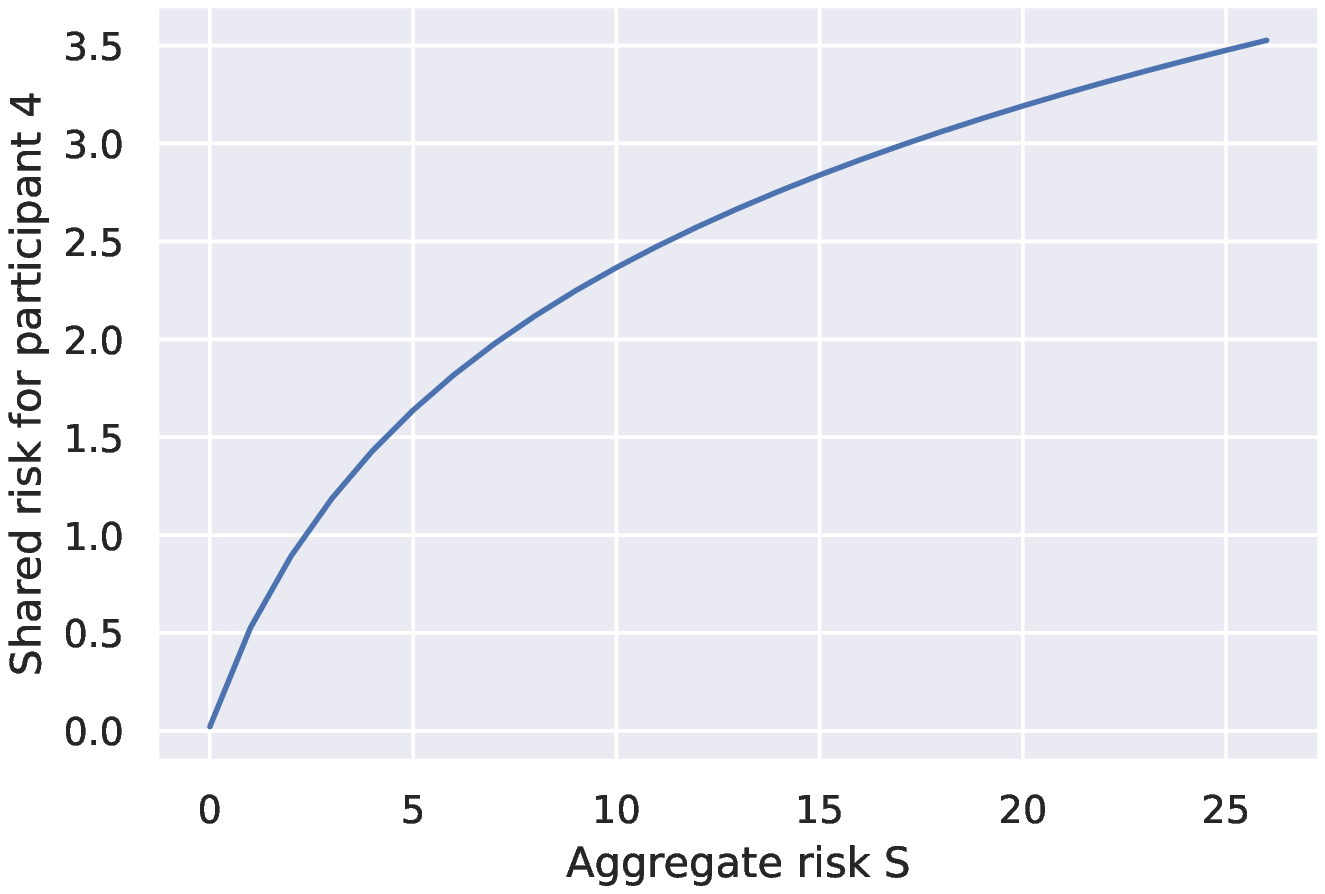}
            \end{minipage}
            \caption{Risk-sharing rules of the four participants for the sensitivity to risk aversion parameters.}
            \label{hsen4}
        \end{figure}

\begin{table}[h!]
            \centering
           
            \begin{tabular}{|c|c|c|c|c|c|c|c|c|}
                \hline
                Participant $i$ & 1 & 2 & 3 & 4\\
                \hline
                
                $\sigma_i$ & 1 & 2 & 3 & 4 \\
                
                \hline
                $\boldsymbol{\alpha}$ & $0.62$ & $0.22$ & $0.10$ & $0.05$\\ 
                \hline                
            \end{tabular}
            \caption{Marginal parameters for the sensitivity to risk aversion parameters.}
            \label{param4bis}
\end{table}
We observe that  the components of the fixed point $\boldsymbol{\alpha}$ depend on participants' risk aversion. In fact, the $\alpha_i$ decreases as the risk aversion coefficient $\sigma_i$ of the participant increases, indicating a lower weight on disutility. The risk-sharing rules $h_i$ ($i=1, \dots, 4$) are represented in Figure \ref{hsen4}. One can note in this figure that for lower values of the aggregate risk $S$, participants with higher relative risk aversion coefficients bear more risk. Conversely, for large realizations of $S$, the least risk-averse participants bear more risk.

\begin{figure}[h!]
       
        \centering
            \begin{minipage}{0.48\linewidth}
                \centering
                \includegraphics[width=.8\linewidth]{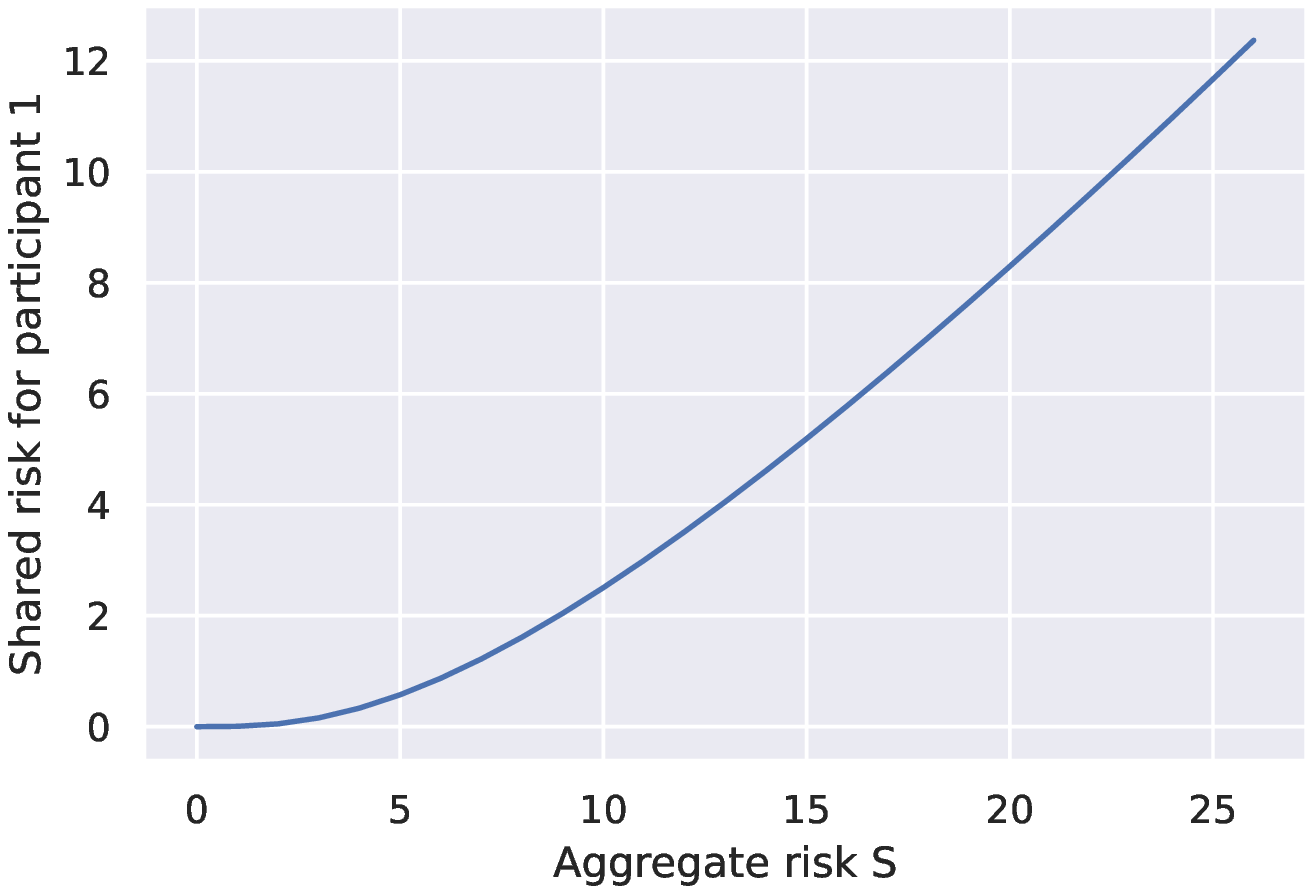}
            \end{minipage}%
            \hspace{4pt}
            \begin{minipage}{0.48\linewidth}
                \centering
                \includegraphics[width=.8\linewidth]{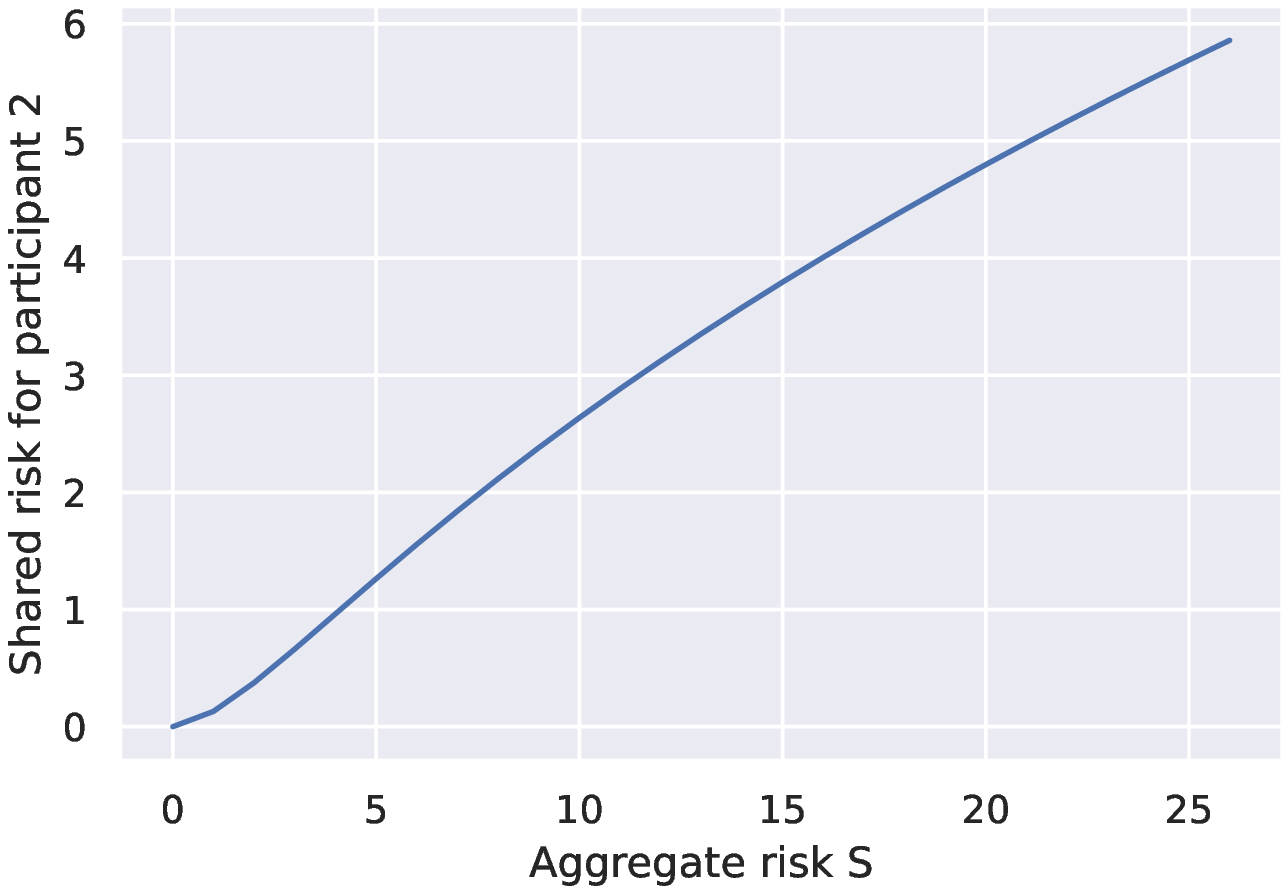}
            \end{minipage}
        \end{figure}

\subsection{Large portfolio of independent compound Poisson distributed losses \label{largp}}
In this section, we provide a numerical example to demonstrate the efficiency of our algorithm in a large portfolio of heterogeneous participants.\\
Consider that the participants use the disutility function of the form $$v_i(s)=\gamma_i\exp\left(\frac{s}{\gamma_i}\right)-s.$$ In this example, $\gamma_i$ is drawn using a uniform distribution on $\{1,2,3,4,5,6,7,8,9,10\}$.\newline
Moreover, it is assumed that the loss $X_i$ for participant $i$ to the risk-sharing pool can be represented as  

\begin{equation}
    X_i=\sum_{k=1}^{N_i}C_{ik} \ \text{with} \ N_i \sim \text{Poisson}(\lambda_i), \quad\forall \ i \ \in \{1, \dots, n \}
\end{equation}
where the claim severities $C_{ik}$ are positive, all these random variables being independent. The random variables $C_{ik}$ are identically distributed (as $C_i$ say) and are assumed to be valued in $\{1, 2, 3, \dots \}$. More specifically, one supposes in this example that $C_i \sim \text{Nbinom}(r_i,q_i)$ where $\text{Nbinom}$ refers to Negative binomial.
This is a reasonable assumption for actuarial applications, where claim severities are expressed as multiples of an appropriate monetary unit (after discretization). We thus explicitly allow for heterogeneity among participants in both loss frequencies $N_i$ and loss severities $C_i$.\\
The application is parametrized as follows:

\begin{itemize}
    \item Consider a pool of $n=1000$ participants;
    \item $N_i \sim \text{Poisson}(\lambda_i)$ where $\lambda_i$ is drawn using an exponential distribution of parameter $10$: $\lambda_i \sim \text{Exp(10)}$ for $i=1, \dots,n$;
    \item $C_i \sim \text{Nbinom}(r_i,q_i)$ where $r_i$ is drawn using an uniform distribution on $\{1,2,3,4,5,6\}$ i.e. $r_i \sim \text{Unif}(\{1,2,3,4,5,6\})$ and $q_i$ is drawn using an uniform distribution on $[0.4,0.5]$ i.e. $q_i \sim \text{Unif}([0.4,0.5])$ for $i=1, \dots,n$;
    \item The average values of parameters are $\bar{\lambda}=0.1$, $\bar{r}=3.5$ and $\bar{q}=0.45$.
\end{itemize}
Table \ref{param10} presents the parameters associated with the first 8 participants.
\begin{table}[h!]
            \centering
           \scalebox{0.8}{%
            \begin{tabular}{|c|c|c|c|c|c|c|c|c|}
                \hline
                Participant $i$ & 1 & 2 & 3 & 4  & 5 & 6 & 7 & 8\\
                \hline
                $\lambda_i$ & 0.090 & 0.430 & 0.045 & 0.003  & 0.009 & 0.041 & 0.029 & 0.147\\

                $q_i$ & 0.419 & 0.493 & 0.480 & 0.401  & 0.491 & 0.441 & 0.430 &  0.418\\

                $r_i$ & 1 & 4 & 4 & 4  & 5 & 2 & 2 & 4\\
      
                $\gamma_i$ & 3 & 2 & 5 & 9  & 8 & 4 & 1 & 9\\
           
                $E\left[X_i\right]$ & 0.125 & 1.767 & 0.194 & 0.018  & 0.045 & 0.104 & 0.076 & 0.816\\
                \hline
                $\boldsymbol{\alpha}$ & $1.8\times10^{-4}$ & $5.3\times10^{-6}$ & $1.9\times10^{-4}$ & $3.9\times10^{-3}$ & $1.4\times10^{-3}$ & $2.9\times10^{-4}$ & $9.6\times10^{-5}$ & $8.0\times10^{-5}$\\ 
                \hline
            \end{tabular}}
            \caption{First 8 participants set of parameters for example in Section \ref{largp}.}
            \label{param10}
        \end{table}
        
\begin{figure}[h!]
        \centering
                \includegraphics[width=.5\linewidth]{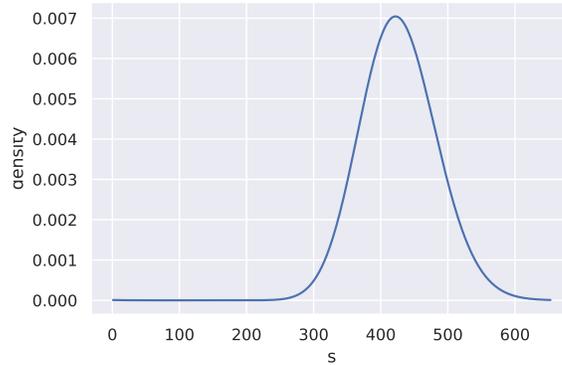}
                \caption{Probability mass function of the aggregated random variable $S$.}
                \label{density}
\end{figure}

        \noindent Figure \ref{density} illustrates the probability mass function of the aggregate random variable $S$.

\begin{figure}[h!]
        \centering
                \includegraphics[width=.6\linewidth]{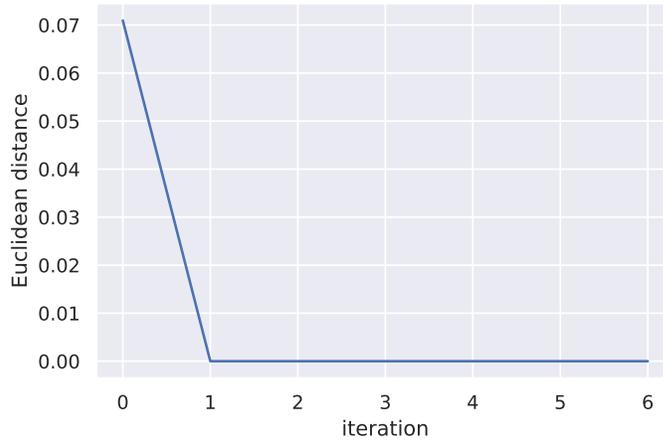}
                \caption{Euclidean distance as a function of the number of iterations.}
                \label{conv}
\end{figure}

\noindent The iterative algorithm, which starts at the point $\alpha^0=(\frac{1}{1000}, \dots, \frac{1}{1000})$, produces a solution in two iterations that satisfies the Euclidean distance constraint \eqref{myfeasib2} up to an error less than $10^{-14}$ (see Figure \ref{conv}). The actuarial fairness constraints in \eqref{formule32} are fulfilled up to machine precision by the design of the algorithm. 

\noindent
In Figure \ref{large} below, the evolution of $h_i$ as a function of the aggregate risk $S$ for the first 8 participants are represented. We observe that the risk-sharing rules are close to the linear risk-sharing rule except for participant 2. The values of the fixed point $\alpha$ for these participants are in Table \ref{param10}. One can see that participant 2 has the lowest disutility weight, which explains the concave shape of his/her risk-sharing rule compared to the others. A more in-depth analysis of participant 2 reveals that he/she is actually the riskiest of the participants in terms of expected risks.

        \begin{figure}[h!]
        \centering
            \begin{minipage}{0.48\linewidth}
                \centering
                \includegraphics[width=.8\linewidth]{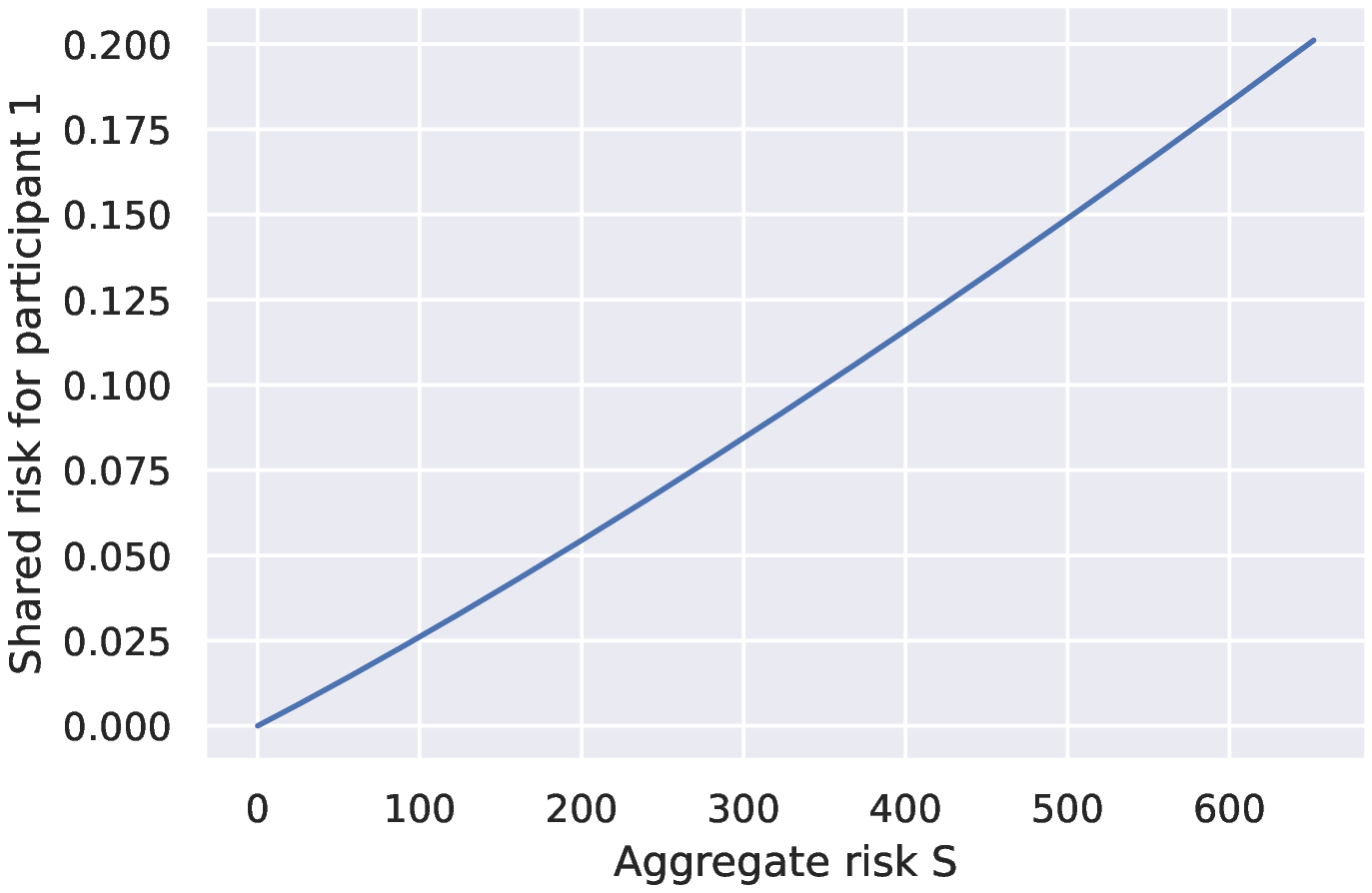}
                
            \end{minipage}%
            \hspace{4pt}
            \begin{minipage}{0.48\linewidth}
                \centering
                \includegraphics[width=.8\linewidth]{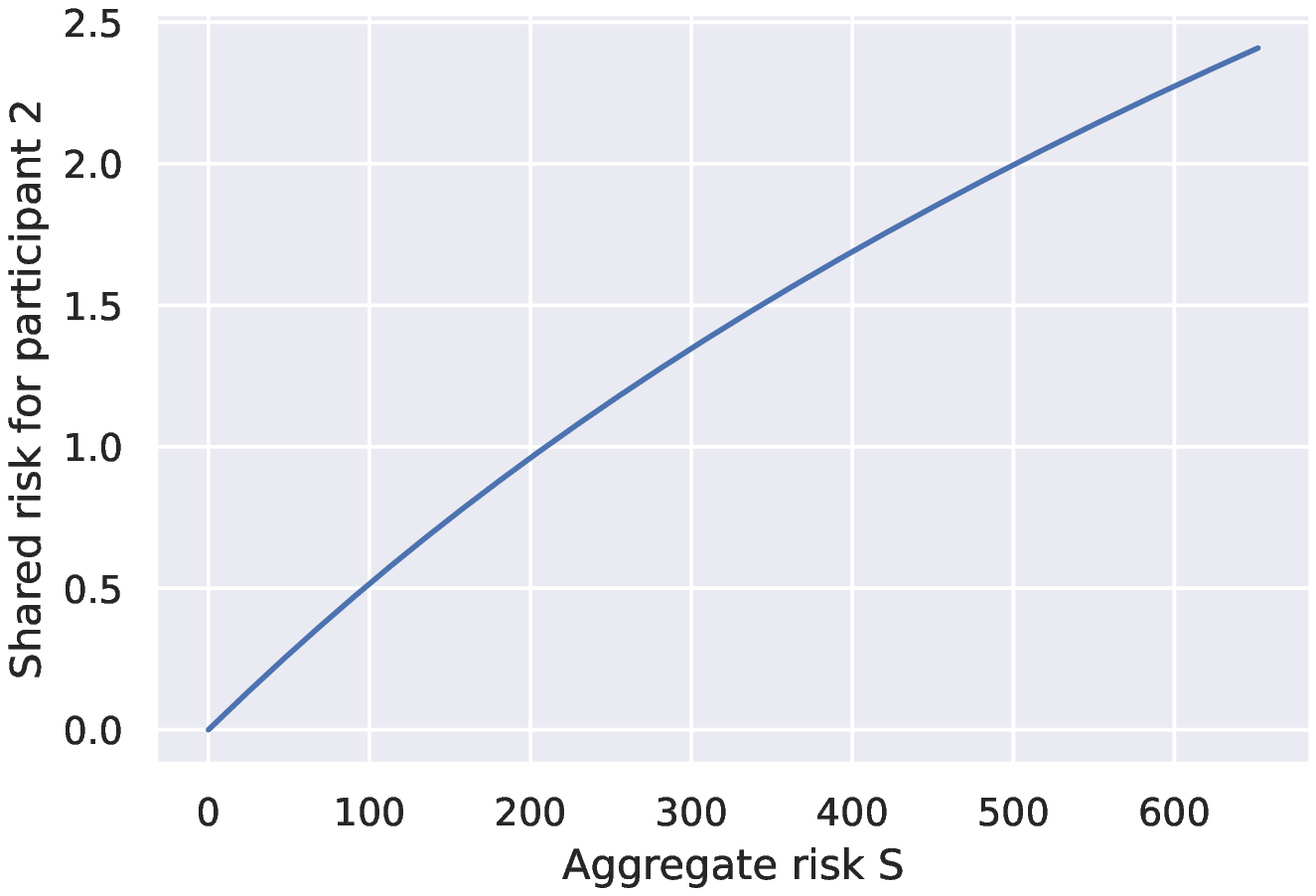}
            \end{minipage}
        \end{figure}

        \begin{figure}[h!]
        \centering
            \begin{minipage}{0.48\linewidth}
                \centering
                \includegraphics[width=.8\linewidth]{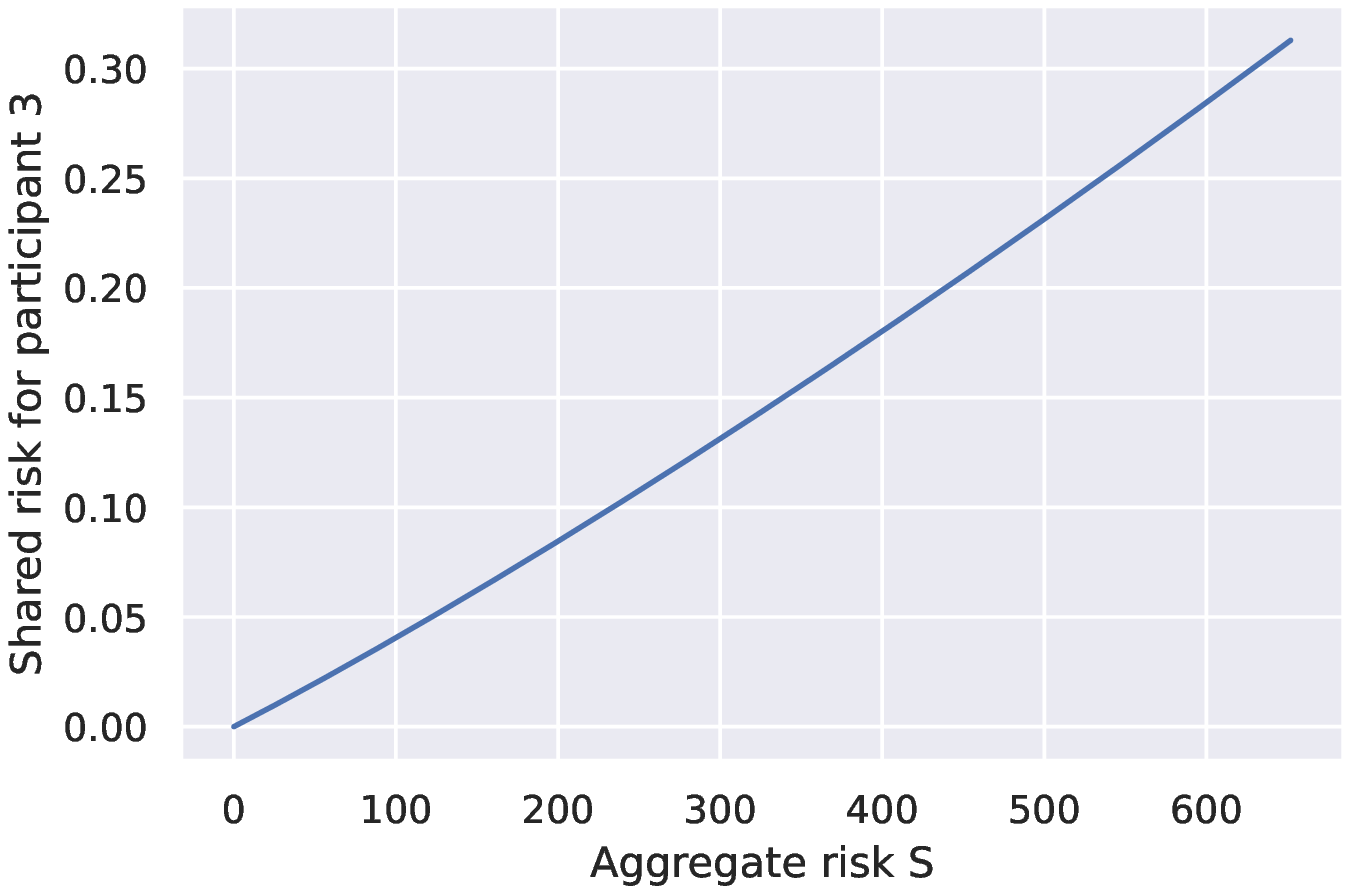}
            \end{minipage}%
            \hspace{4pt}
            \begin{minipage}{0.48\linewidth}
                \centering
                \includegraphics[width=.8\linewidth]{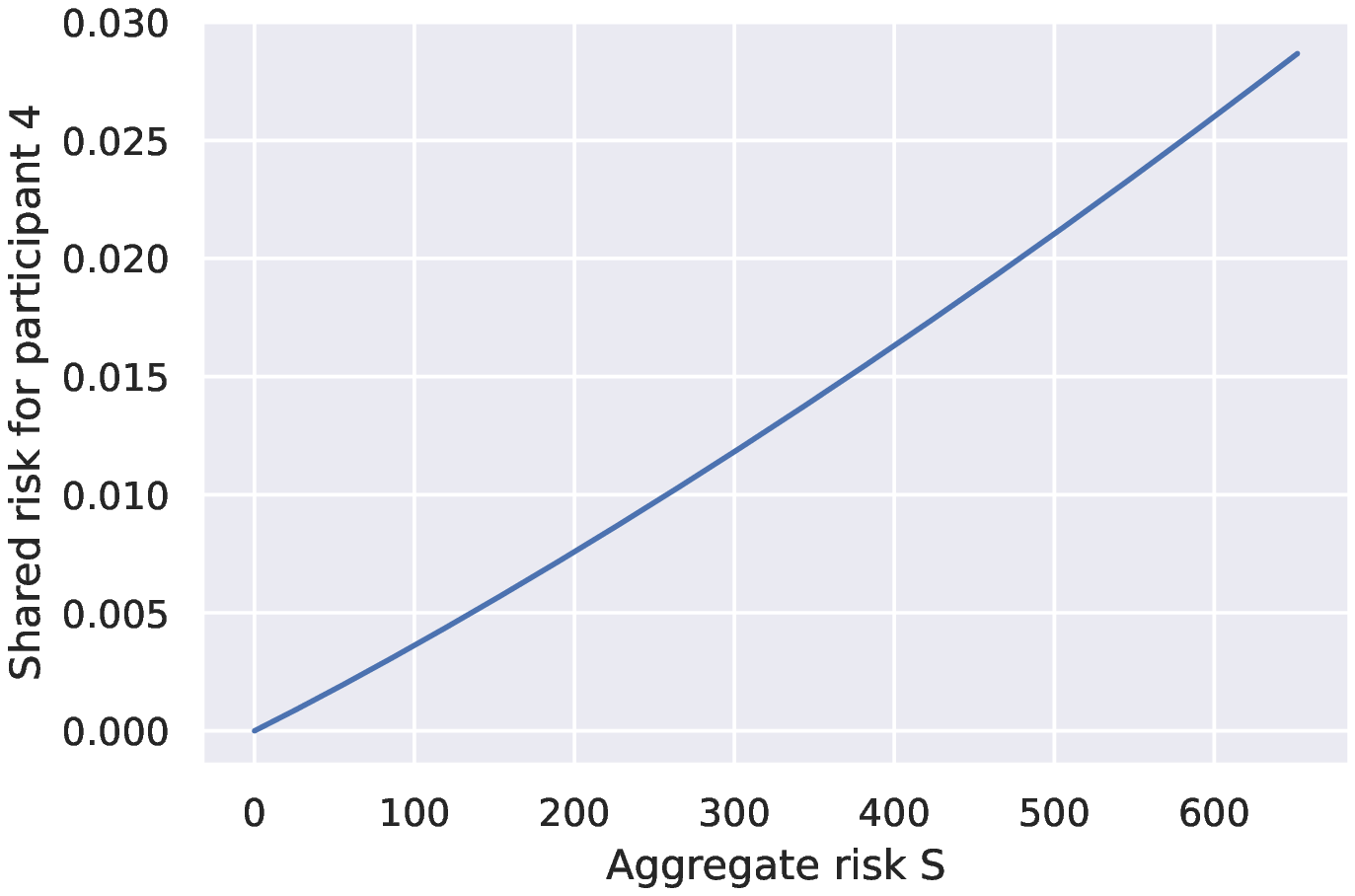}
            \end{minipage}
            
        \end{figure}

        \begin{figure}[h!]
        
        \centering
            \begin{minipage}{0.48\linewidth}
                \centering
                \includegraphics[width=.8\linewidth]{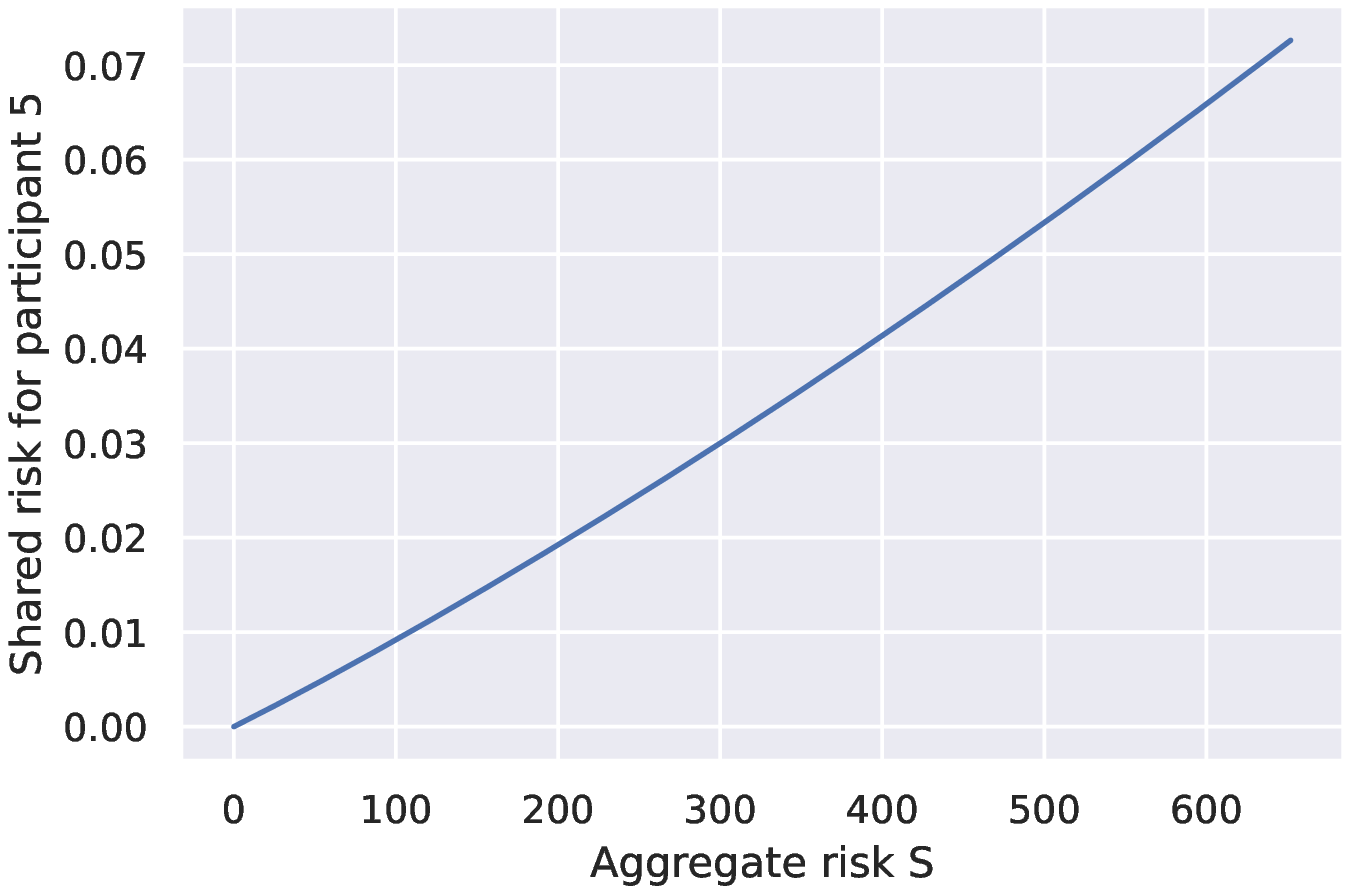}
            \end{minipage}%
            \hspace{4pt}
            \begin{minipage}{0.48\linewidth}
                \centering
                \includegraphics[width=.8\linewidth]{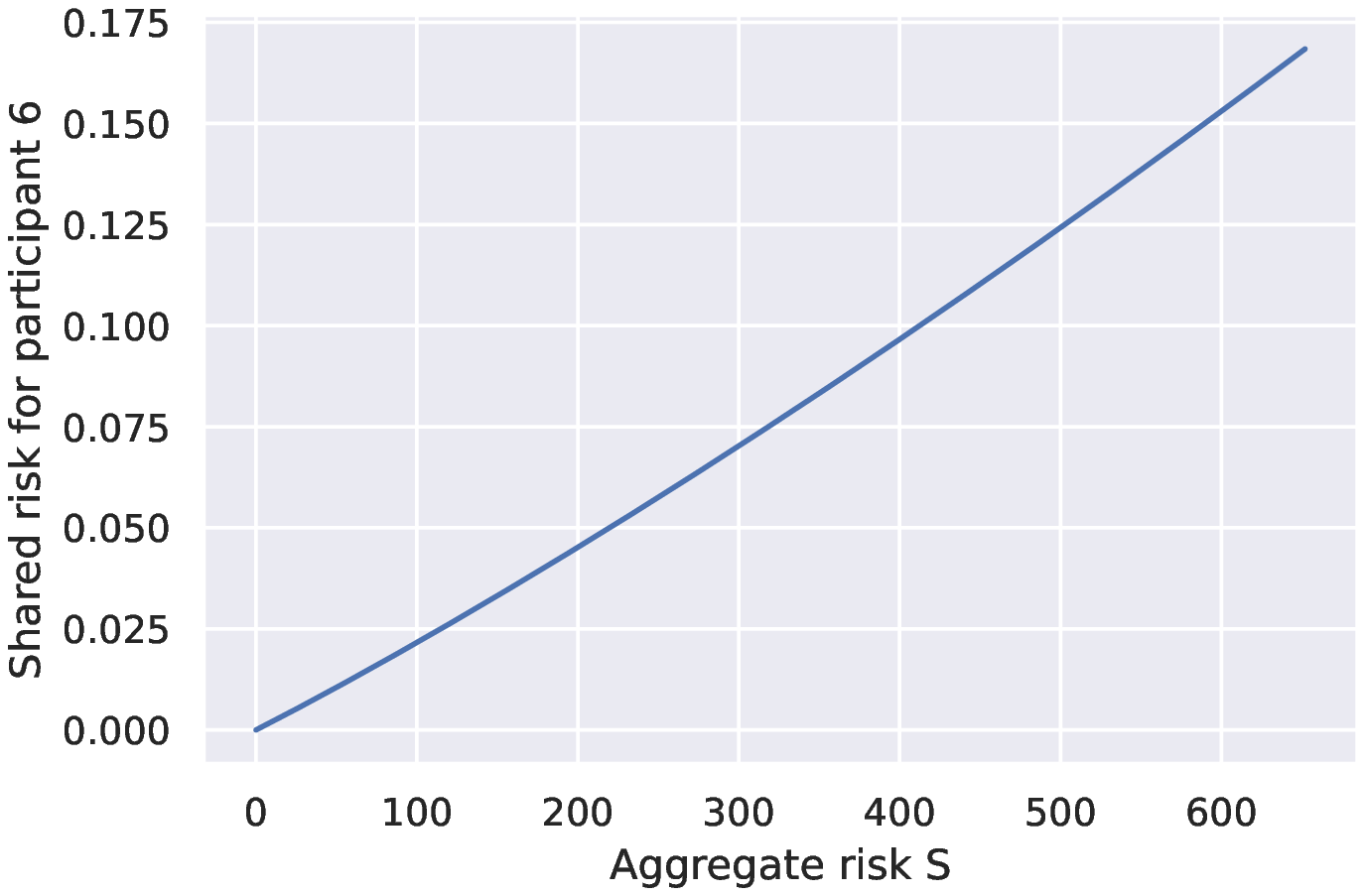}
            \end{minipage}
        \end{figure}

        \begin{figure}[h!]
        \centering
            \begin{minipage}{0.48\linewidth}
                \centering
                \includegraphics[width=.8\linewidth]{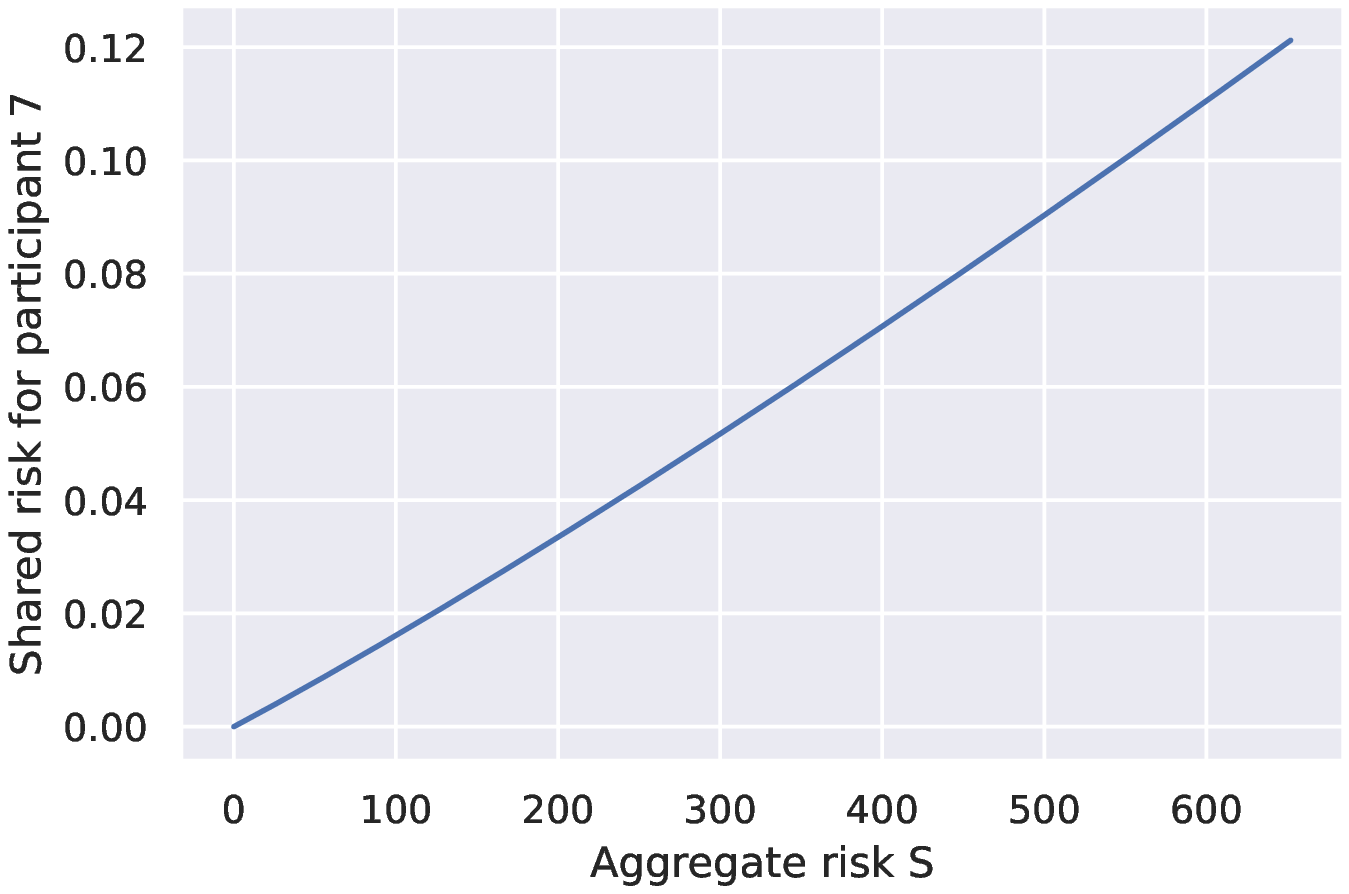}
            \end{minipage}%
            \hspace{4pt}
            \begin{minipage}{0.48\linewidth}
                \centering
                \includegraphics[width=.8\linewidth]{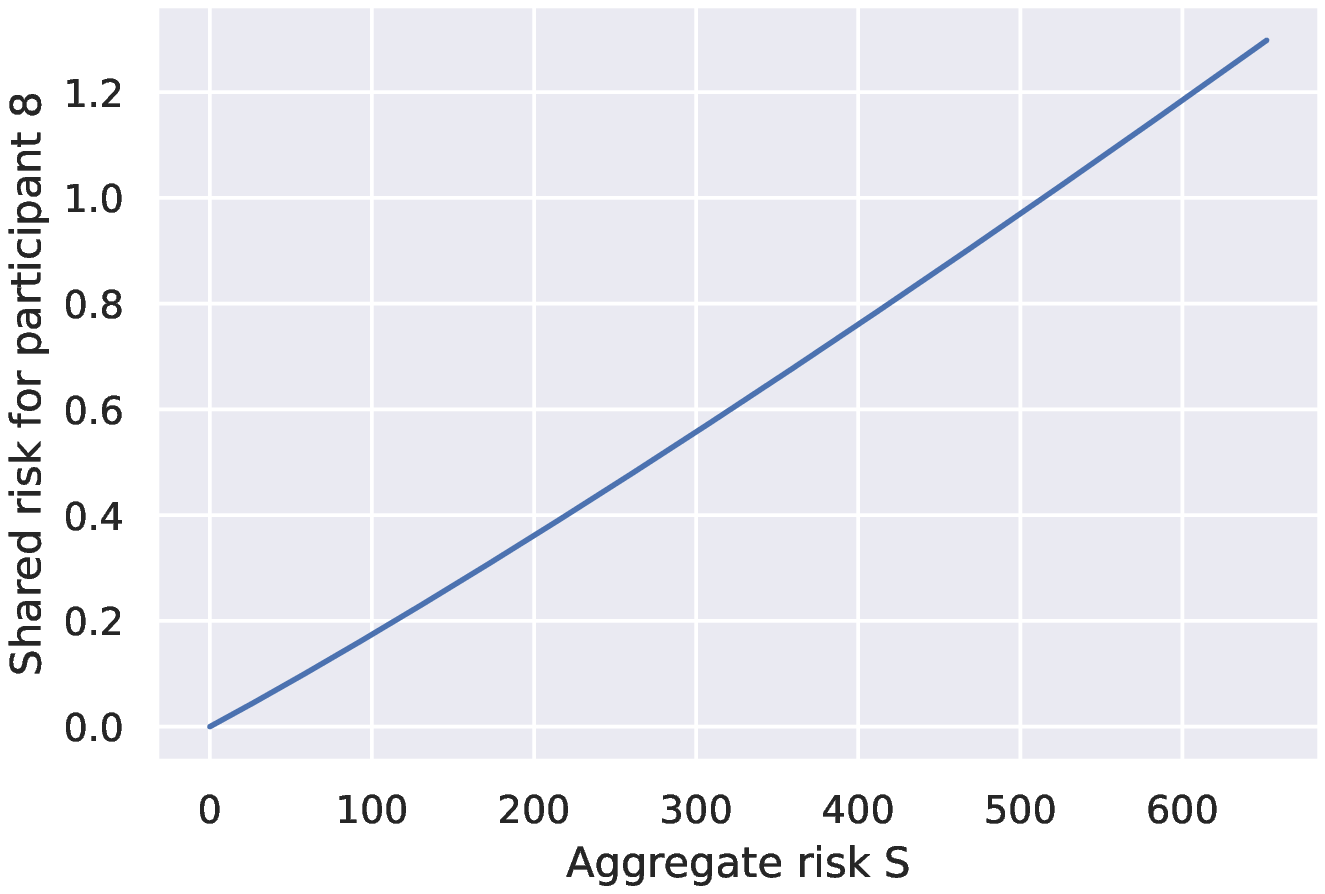}
            \end{minipage}
            \caption{Risk-sharing rules of the eight participants for example in Section \ref{largp}.}
            \label{large}
        \end{figure}

\section{Convex order comparative statics of AFPO risk-sharing allocations \label{cxd}}
This section  investigates the property that for two pools associated with two aggregate risks $S$ and $\tilde{S}$ that are comparable in the convex order sense, the AFPO risk-sharing rules associated with these two pools are also comparable in this order.

\subsection{Overview of convex order}

\begin{definition}
Let $X$ and $Y$ be two random variables such that
\begin{equation}
  \mathbb{E}[g(X)] \leq \mathbb{E}[g(Y)] \quad \text { for all convex functions } \quad g: \mathbb{R} \rightarrow \mathbb{R}  
\end{equation}
provided the expectations exist. Then $X$ is said to be smaller than $Y$ in the convex order (denoted as $X \preceq_{\mathrm{CX}} Y$ ). $X$ is said to be strictly smaller than $Y$ in convex order, which is denoted as $X \prec_{\mathrm{CX}} Y$, if $X \preceq_{\mathrm{CX}} Y$ holds true and $X$ and $Y$ are not identically distributed.
The stochastic inequality $X \preceq_{\mathrm{CX}} Y$ intuitively means that $X$ and $Y$ have the same magnitude (as $\mathbb{E}[X]=\mathbb{E}[Y]$ holds) but that $Y$ is more variable than $X$. For instance, the variance of $Y$ is larger than the variance of $X$. For a thorough description of the convex order and its applications in insurance studies, we refer the reader to \cite{denuit_actuarial_2005}. A general reference about stochastic order relations is \cite{shaked_stochastic_2007}.
\label{defcx}
\end{definition}
Establishing the relation $X \preceq_{\mathrm{CX}} Y$ between two random variables $X$ and $Y$ can be very useful. However, it is sometimes not clear how to verify that $X \preceq_{\mathrm{CX}} Y$ given the two random variables and their distributions. A sufficient condition that leads to the convex order is the \emph{Karlin-Novikof cutting theorem}. We present below a version of this theorem from \cite{shaked_stochastic_2007}.

\begin{theorem}
    \emph{Let $X$ and $Y$ be two random variables with equal means, density functions $f$ and $g$, distribution functions $F$ and $G$, and survival functions $\bar{F}$ and $\bar{G}$, respectively. Let us note by $\mathfrak{S}^{-}(\ell)$ the number of sign changes of the function $\ell$. Then $X \preceq_{\mathrm{CX}} Y$ if any of the following conditions hold:}
        $$
        \begin{aligned}
        &\mathfrak{S}^{-}(g-f)=2 \text { \emph{and the sign sequence is} }+,-,+\text {, } \\
        &\mathfrak{S}^{-}(\bar{F}-\bar{G})=1 \text { \emph{and the sign sequence is} }+,-\text {, } \\
        &\mathfrak{S}^{-}(G-F)=1 \quad \text {\emph{and the sign sequence is} }+,- \text{.}\\
        &
        \end{aligned}
        $$
\end{theorem}
\noindent This characterization of convex order is used in this paper to prove Proposition \ref{premium}. 

\subsection{Convex order comparative statics condition}
Here, we present the conjecture to be proven in order to validate the convex order comparative statics of AFPO risk-sharing allocations. 

    \begin{conjecture}
           \emph{Consider two risk sharing schemes $(h_1(S), \dots, h_n(S))$ and $(\tilde{h}_1(\tilde{S}), \dots, \tilde{h}_n(\tilde{S}))$ associated with $S=\sum_{i=1}^{n}X_{i}$ and $\tilde{S}=\sum_{i=1}^{n}\Tilde{X}_{i}$ s.t.  $\mathbb{E}[X_i]=\mathbb{E}[\tilde{X}_i]$ for all $i \ \in \ \{1, \dots, n\}$.}\bigskip
                
           \emph{If $S \preceq_{\mathrm{CX}} \tilde{S}$ then $h_i(S) \preceq_{\mathrm{CX}} \tilde{h}_i(\tilde{S})$ for all $i \ \in \ \{1, \dots, n\}.$}
    \end{conjecture}\bigskip
    
\noindent We were able to demonstrate this conjecture in the context of the example below.
    \begin{proposition}
           \emph{Consider two risk sharing schemes $(h_1(S), h_2(S))$ and $(\tilde{h}_1(\tilde{S}), \tilde{h}_2(\tilde{S}))$ associated with $S=\sum_{i=1}^{2}X_{i}$ and $\tilde{S}=\sum_{i=1}^{2}\Tilde{X}_{i}$ s.t.  $\mathbb{E}[X_i]=\mathbb{E}[\tilde{X}_i]$ for all $i \ \in \ \{1, 2\}$.}\\
            \emph{In each pool, risk aversion is characterized by the following power disutility function:
            $$v_i(s)=\frac{s^{1+\sigma_i}}{1+\sigma_i},$$ 
             with $\sigma_2=2 \sigma_1$ and $\tilde{\sigma}_2=2 \tilde{\sigma}_1.$ 
            We assume that $X_i \ \sim \ \Gamma(\mu_i, \lambda)$ with $X_1 \indep X_2$ and $\tilde{X}_i \ \sim \ \Gamma(\tilde{\mu}_i, \tilde{\lambda})$ with $\tilde{X}_1 \indep \tilde{X}_2$. In that case, $S \ \sim \ \Gamma(\mu, \lambda)$ and $\tilde{S} \ \sim \ \Gamma(\tilde{\mu}, \tilde{\lambda})$ where $\mu=\mu_1+\mu_2$ and $\tilde{\mu}=\tilde{\mu}_1+\tilde{\mu}_2$.\newline
            We choose $\mu_1, \mu_2, \tilde{\mu}_1,\tilde{\mu}_2, \lambda, \tilde{\lambda}$ such that: $\mu \geq \tilde{\mu}$ and $\frac{\mu}{\lambda}=\frac{\tilde{\mu}}{\tilde{\lambda}}.$ These conditions on the parameters lead to $S \preceq_{\mathrm{CX}} \tilde{S}.$ The following result holds}\bigskip
            
           \emph{ $$h_i(S) \preceq_{\mathrm{CX}} \tilde{h}_i(\tilde{S}) \ \text{for all} \ i \ \in \ \{1, 2\}.$$}
           \label{premium}
    \end{proposition}
The proof of Proposition \ref{premium} is given in Appendix \ref{pre1}.
In the following section, a numerical illustration of this conjecture is given.
\subsection{Numerical illustration}
\begin{figure}[h!]
        \centering
            \begin{minipage}{0.3\linewidth}
                \centering
                \includegraphics[width=.9\linewidth]{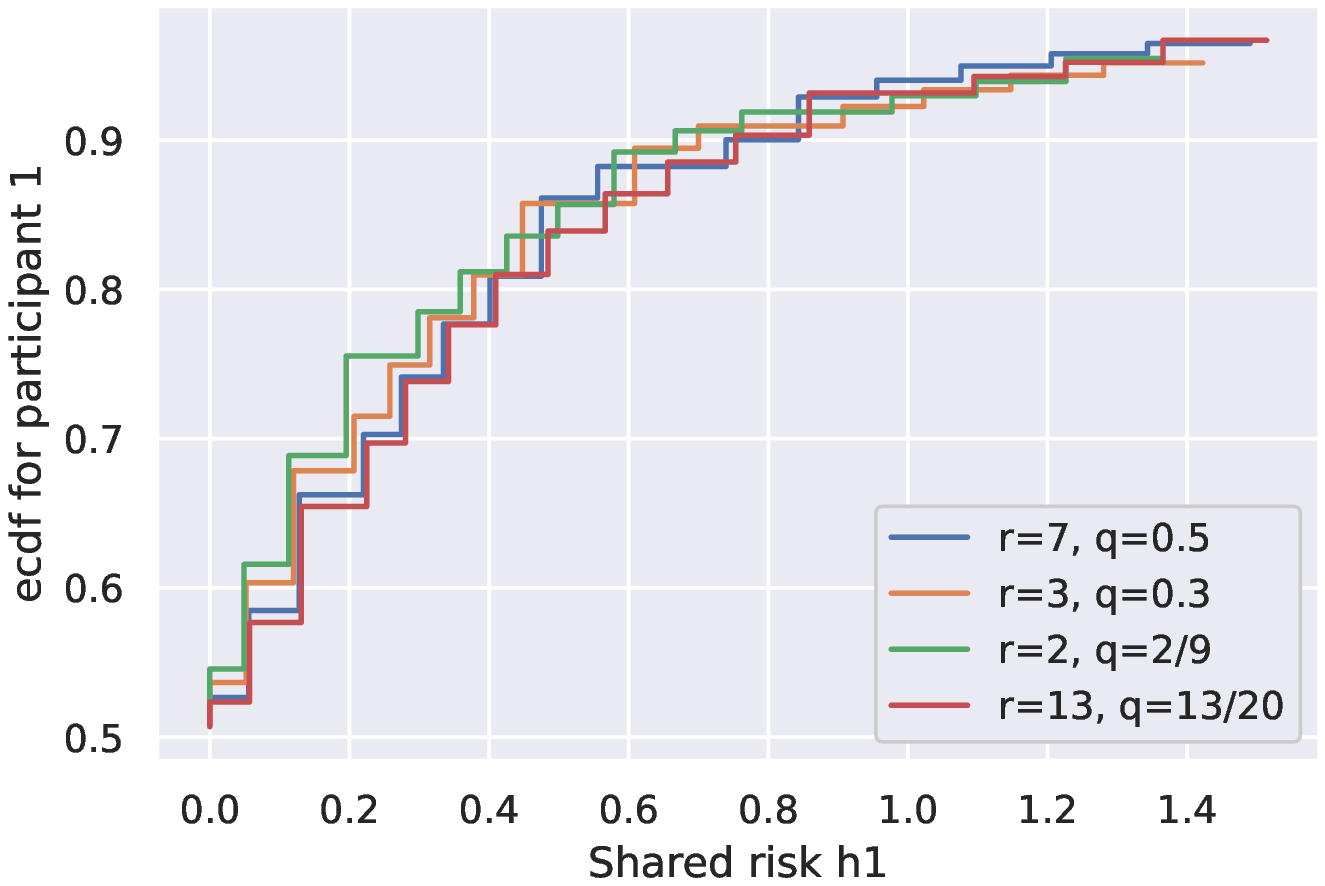}
            \end{minipage}%
            \hspace{2pt}
            \begin{minipage}{0.3\linewidth}
                \centering
                \includegraphics[width=.9\linewidth]{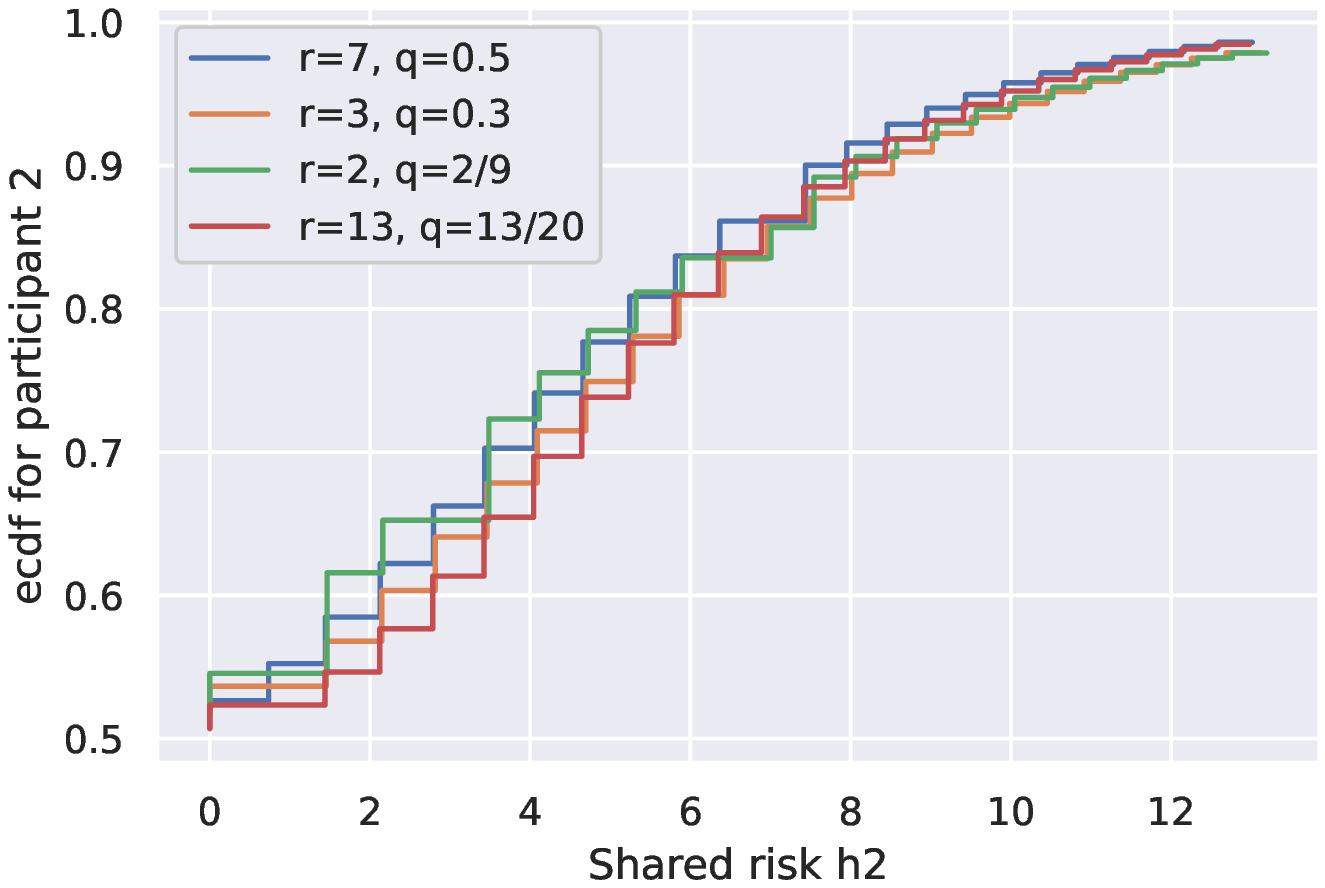}
            \end{minipage}
            \hspace{2pt}
            \begin{minipage}{0.3\linewidth}
                \centering
                \includegraphics[width=.9\linewidth]{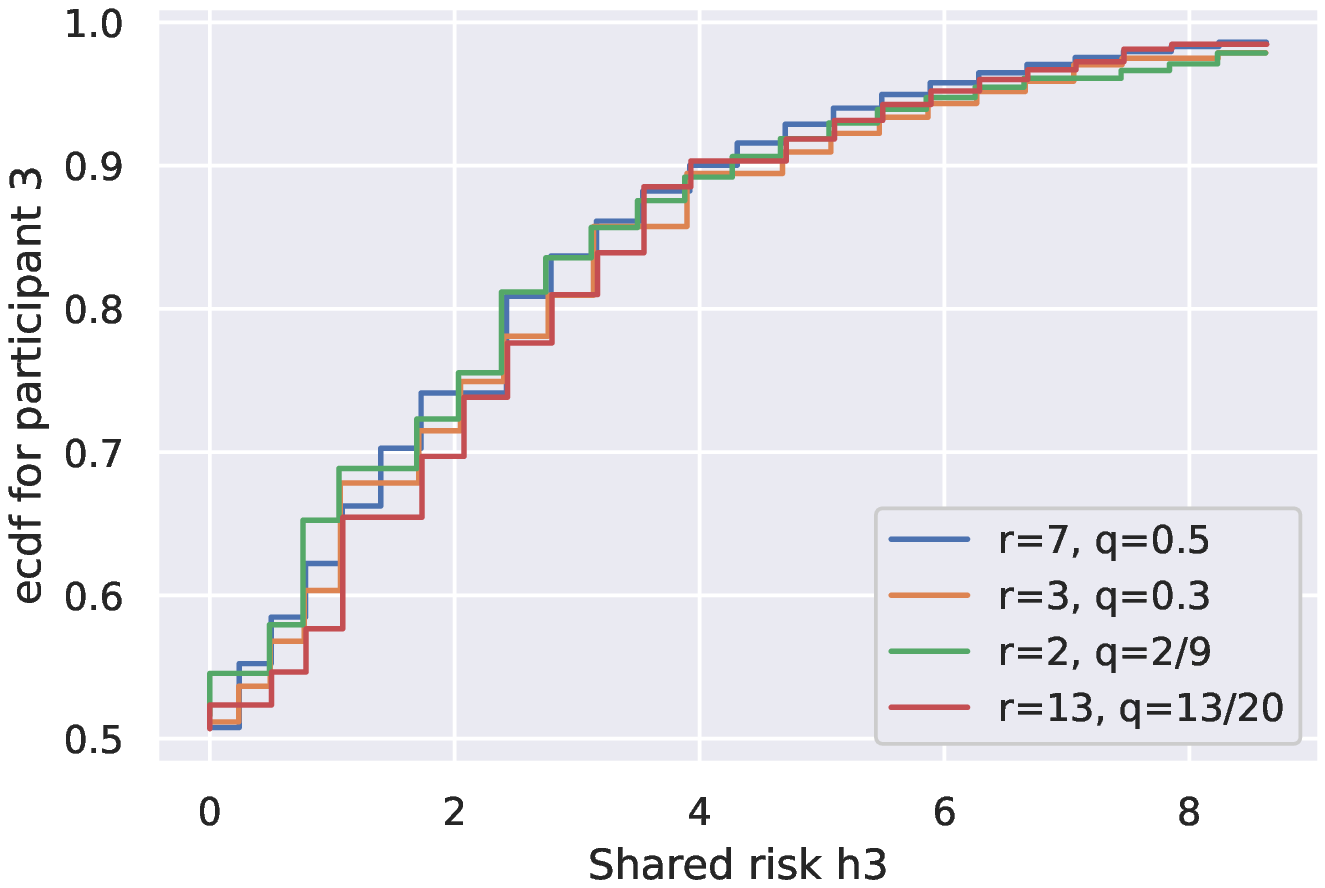}
            \end{minipage}
            \caption{Cdfs in each case for the three participants.}
            \label{cdf}
        \end{figure}

In this part, we want to illustrate numerically the convex order property of AFPO risk-sharing rules. This property studies the behavior of AFPO risk-sharing rules when replacing the original loss $X_i$ of a given participant $i$ by a new loss $\tilde{X_i}$ which is “of the same size” (in the sense that expectations are equal) but which is “more or less variable”. The losses of the others participants remain unchanged. In this case, the mean of the aggregated loss $S$ remains unchanged but it is “more or less variable”. We realize this study by considering three participants and assume that the losses follow a compound Poisson distribution as above. The application is parametrized as follows:
\begin{itemize}
    \item for the baseline, we compute the risk-sharing rule with $(\lambda_1,\lambda_2,\lambda_3) = (0.1,0.4,0.2)$, $(q_1,q_2,q_3) = (0.42,0.5,0.45)$, $(r_1,r_2,r_3) = (2,7,6)$;
    \item case 1: $q_2=0.3$ and $r_2=3$. This corresponds to the same mean but more variable w.r.t the baseline;
    \item case 2: $q_2=2/9$ and $r_2=2$. This corresponds to the same mean but more variable w.r.t to use case 1;
    \item  case 3: $q_2=13/20$ and $r_2=13$. This corresponds to the same mean but less variable w.r.t the baseline.
\end{itemize}
In order to demonstrate numerically the convex order property of AFPO risk-sharing rules, we represent in Figure \ref{cdf} the empirical cumulative distribution functions (cdfs) of the three risk-sharing rules. By applying \emph{Karling-Novikoff} cutting theorem, it is enough to show that the curves of each use-case cross once the curve of the baseline. 
Table \ref{cdf1} below crosses the cdfs as a function of $q$ and indicates whether these cdfs intersect once. We indicate, if applicable, the S value of the intersection point. It is important to note that this table is the same for the three participants.  

\begin{table}[h!]
            \centering
           
            \begin{tabular}{|c|c|c|c|c|}
                \hline
                Cdf of $h_i(S)$ & "r=7, q=0.5" & "r=3, q=0.3" & "r=2, q=2/9" & "r=13, q=13/20"\\
                \hline
                "r=7, q=0.5" & - & $S=9$ & $S=9$ & $S=8$ \\

                "r=3, q=0.3" &  & - & $S=10$ & $S=8$ \\

                "r=2, q=2/9" &  &  & - & $S=9$ \\
      
                "r=13, q=13/20" &  &  &  & - \\
                \hline
            \end{tabular}
            \caption{Intersection between cdfs of $h_i(S)$.}
            \label{cdf1}
        \end{table}

\section{Conclusion \label{sect5}}
In this paper, we have proposed an iterative algorithm for \emph{actuarially fair Pareto optimal} (AFPO) risk-sharing rules in \emph{'non-olet'} risk-sharing system. A one-to-one correspondence between AFPO risk-sharing rules and fixed points of a composite iteration mapping is established. This allowed us to show the existence and uniqueness of solutions and the global convergence of the corresponding iteration map. We have considered a finite number of \emph{von Neumann Morgenstern} participants where each of them used a convex disutility function of taking on a risk.
The present paper offers practitioners a fast numerical method for computing general risk-sharing rules, for a single-period risk-sharing system. Nevertheless, a multi-period risk-sharing is more convenient for practitioners: the next step in this span of research is to develop a model along the lines of the present paper but for a multi-period case. Furthermore, the numerical procedure allows for straightforward testing of the proposed conjecture for the convex order, but further development is needed in future work.

\section*{Acknowledgement}
I am grateful to my Ph.D. supervisors, Caroline Hillairet and Christian Robert, who made countless efforts to review and correct this paper. My acknowledgment also goes to Christopher Blier-Wong for his useful comments which helped to improve the text.

 \appendix 

\section*{Appendix}

\section{Proof of solutions \eqref{h2_nonl} and \eqref{h1_nonl} \label{A15}}
One finds $J(.)$, $\alpha_1$ and $\alpha_2$ by solving

\begin{itemize}
    \item Feasibility condition: \begin{equation}
        I_1\left(\frac{J(s)}{\alpha_1}\right)+I_2\left(\frac{J(s)}{\alpha_2}\right)=s .
        \label{fair}
    \end{equation}
    \item Fairness for participant $1$:
    \begin{equation}
        \mathbb{E}\left[I_1\left(\frac{J\left(S\right)}{\alpha_1}\right)\right]=\mathbb{E}\left[X_1\right] .
        \label{fair1}
    \end{equation}
    \item Fairness for participant $2$: 
    \begin{equation}
        \mathbb{E}\left[I_2\left(\frac{J\left(S\right)}{\alpha_2}\right)\right]=\mathbb{E}\left[X_2\right] .
        \label{fair2}
    \end{equation}

\end{itemize}
The three equations are actually reduced to only two equations. Indeed if the two equations \eqref{fair} and \eqref{fair2} are satisfied then \eqref{fair1} is also satisfied. Therefore, we will look for $J(.)$, $\alpha_1$ and $\alpha_2$ by solving only equations \eqref{fair} and \eqref{fair2}.\newline
We first determine $J(.)$ as a function of $s$, $\alpha_1$ and $\alpha_2$ by solving \eqref{fair}.
        $$
        \begin{aligned}
        &I_1\left(\frac{J(s)}{\alpha_1}\right)+I_2\left(\frac{J(s)}{\alpha_2}\right)=s\\
        &\Rightarrow\left[\frac{J(s)}{\alpha_1}\right]^{\frac{1}{\sigma_1}}+\left[\frac{J(s)}{\alpha_2}\right]^{\frac{1}{\sigma_2}}=s\\
        &\Rightarrow\left[J(s)\right]^{\frac{1}{\sigma_2}}\left[\frac{\left[J(s)\right]^{\left(\frac{1}{\sigma_1}-\frac{1}{\sigma_2}\right)}}{\left(\alpha_1\right)^{1 / \sigma_1}}+\frac{1}{\left(\alpha_2\right)^{1 / \sigma_2}}\right]=s.\\
        &
        \end{aligned}
        $$
Let $H(s)=\left[J(s)\right]^{\frac{1}{\sigma_2}}$ then one has:
        $$
        \begin{aligned}
        &\frac{[H(s)]^2}{\left(\alpha_1\right)^{1 / \sigma_1}}+\frac{1}{\left(\alpha_2\right)^{1 / 2 \sigma_1}} H(s)-s=0\\
        &\Leftrightarrow\left(\alpha_1\right)^{-1 / \sigma_1}[H(s)]^2+\left(\alpha_2\right)^{-1 / 2 \sigma_1} H(s)-s=0.\\
        &
        \end{aligned}
        $$
The discriminant $\Delta$ of this second-degree equation is given by:
$$\Delta(s)=\left(\alpha_2\right)^{-1 / \sigma_1}+4 s\left(\alpha_1\right)^{-1 / \sigma_1}.$$
Knowing that the function $J(.)$ is positive, the appropriate solution is the following:
$$ \begin{aligned} &J\left(s\right)=[H(s)]^{2 \sigma_1}  &\Rightarrow J\left(s\right)=\frac{\left[\sqrt{\Delta(s)}-\left(\alpha_2\right)^{-1 / 2 \sigma_1}\right]^{2 \sigma_1}}{2^{2 \sigma_1}\left(\alpha_1\right)^{-2}}.  \end{aligned} $$
We finally deduce the expression of $J(.)$ as a function of $\alpha_1$ and $\alpha_2$ as follows: 
\begin{equation}
\boxed{
    J\left(s\right)=\frac{\left(\alpha_1\right)^2}{4^{\sigma_1}}\left(\sqrt{\Delta(s)}-\left(\alpha_2\right)^{-1 / 2 \sigma_1}\right)^{2 \sigma_1}.
    \label{newj}
    }
\end{equation}
One can now replace $J\left(S\right)$ in the fairness equation \eqref{fair2} using \eqref{newj}
$$\mathbb{E}\left[I_2\left(\frac{J\left(S, \alpha\right)}{\alpha_2}\right)\right]=\mathbb{E}\left[X_2\right]$$ or $I_2\left(\frac{J\left(S, \alpha\right)}{\alpha_2}\right)=\left[\frac{J\left(S, \alpha\right)}{\alpha_2}\right]^{1 / \sigma_2}=\frac{H(S)}{\left(\alpha_2\right)^{1 / 2 \sigma_1}}$ $=\frac{\left(\alpha_1\right)^{1 / \sigma_1}}{2\left(\alpha_2\right)^{1 / 2 \sigma_1}}\left(\sqrt{\Delta(S)}-\left(\alpha_2\right)^{-1 / 2 \sigma_1}\right)$ with 
$$\quad \Delta(S)=4 S\left(\alpha_1\right)^{-1 / \sigma_1}+\left(\alpha_2\right)^{-1 / \sigma_1}.$$ 
Let $\beta=\frac{\mathbb{E}\left[X_2\right]}{\mathbb{E}\left[S\right]}$ be the proportion of risk carried by participant $2$.
        $$
        \begin{aligned}
        &\mathbb{E}\left[X_2\right]=\beta\mathbb{E}\left[S\right]=\frac{1}{\left(\alpha_2\right)^{1 / 2 \sigma_1}}\left[\frac{\left(\alpha_1\right)^{1 / \sigma_1}}{2}\left(\mathbb{E}\left[ \sqrt{\Delta(S)}\right]-\left(\alpha_2\right)^{-1 / 2 \sigma_1}\right)\right]\\
        &\beta\mathbb{E}\left[S\right]=\frac{\left(\alpha_1\right)^{1 / \sigma_1} \left(\alpha_2\right)^{-1 / 2 \sigma_1}}{2}\mathbb{E} \left[ \sqrt{\Delta(S)}\right]-\frac{\left(\alpha_1\right)^{1 / \sigma_1} \left(\alpha_2\right)^{-1 / 2 \sigma_1} \left(\alpha_2\right)^{-1 / 2 \sigma_1}}{2}\\
        &\beta\mathbb{E}\left[S\right]=\frac{\left(\alpha_1\right)^{1 / \sigma_1} \left(\alpha_2\right)^{-1 / 2 \sigma_1}}{2}\mathbb{E}\left[\sqrt{4 S\left(\alpha_1\right)^{-1 / \sigma_1}+\left(\alpha_2\right)^{-1 / \sigma_1}}\right]-\frac{\left(\alpha_1\right)^{1 / \sigma_1} \left(\alpha_2\right)^{-1 / \sigma_1}}{2}\\
        &\beta\mathbb{E}\left[S\right]=\mathbb{E}\left[\sqrt{ \frac{\left(\alpha_1\right)^{2 / \sigma_1} \left(\alpha_2\right)^{-1 /  \sigma_1}}{4} \times 4 S\left(\alpha_1\right)^{-1 / \sigma_1}+ \frac{\left(\alpha_1\right)^{2 / \sigma_1} \left(\alpha_2\right)^{-1 /  \sigma_1}}{4} \times \left(\alpha_2\right)^{-1 / \sigma_1}}\right]-\frac{\left(\alpha_1\right)^{1 / \sigma_1} \left(\alpha_2\right)^{-1 / \sigma_1}}{2}\\
        &\beta\mathbb{E}\left[S\right]=\mathbb{E}\left[\sqrt{\left(\alpha_1\right)^{1 / \sigma_1} \left(\alpha_2\right)^{-1 /  \sigma_1} S + \frac{\left(\alpha_1\right)^{2 / \sigma_1} \left(\alpha_2\right)^{-2 /  \sigma_1}}{4} }\right]-\frac{\left(\alpha_1\right)^{1 / \sigma_1} \left(\alpha_2\right)^{-1 / \sigma_1}}{2}.\\
        &
        \end{aligned}
        $$
Setting $a=\left(\frac{\alpha_1}{\alpha_2}\right)^{1 / \sigma_1}$, one has
\begin{equation}
\boxed{
 \beta\mathbb{E}\left[S\right]=\mathbb{E}\left[\sqrt{a S + \left(\frac{a}{2}\right)^{2} }-\frac{a}{2}\right]. 
 \label{fixpoint}
 }
\end{equation}
Finding $a$ leads to identifying the fixed point on the unit simplex as follows:
$$\tilde{\alpha_1} =\frac{\alpha_1}{\alpha_1+\alpha_2}=\frac{a^{\sigma_1} \alpha_2}{a^{\sigma_1} \alpha_2 + \alpha_2}=\frac{a^{\sigma_1}}{1+a^{\sigma_1}} \ \text{and} \ \tilde{\alpha_2}=1-\tilde{\alpha_1}= \frac{1}{1+a^{\sigma_1}}$$

$$
   \Rightarrow\boxed{
    \tilde{\alpha_1}=\frac{a^{\sigma_1}}{1+a^{\sigma_1}}
    } \ \text{and} \quad \boxed{
    \tilde{\alpha_2}= \frac{1}{1+a^{\sigma_1}}
    .}
$$
Equation \eqref{fixpoint} has a solution thanks to the intermediate value theorem. Indeed we have that $ \mathbb{E} \sqrt{a S + \left(\frac{a}{2}\right)^{2} }-\frac{a}{2}$ tends to $0$ when $a$ tends to $0$ and tends to $\mathbb{E}\left[S\right]$ when $a$ tends to $\infty$. Since $0 < \beta\mathbb{E}\left[S\right]< \mathbb{E}\left[S\right]$ then there exists an $a$ that satisfies the equation.\newline
$\lim\limits_{a \to \infty}$ $ \mathbb{E}\left[\sqrt{a S + \left(\frac{a}{2}\right)^{2} }-\frac{a}{2}\right]$= $\lim\limits_{a \to \infty}$ $\mathbb{E}\left[\frac{a S}{\sqrt{a S + \left(\frac{a}{2}\right)^{2} }+\frac{a}{2}}\right]$=$\lim\limits_{a \to \infty}$ $\mathbb{E}\left[\frac{S}{\sqrt{\frac{S}{a} + \frac{1}{4} }+\frac{1}{2}}\right]$.
Applying the dominated convergence theorem with the assumption that the distribution $S$ has a finite mean we have

$\lim\limits_{a \to \infty}$ $ \mathbb{E}\left[\sqrt{a S + \left(\frac{a}{2}\right)^{2} }-\frac{a}{2}\right]$ = $\mathbb{E}\left[\frac{S}{\sqrt{\frac{1}{4}}+\frac{1}{2}}\right]=\mathbb{E}\left[S\right].$ \newline
For uniqueness, let us consider the function $f(a)=\mathbb{E}\left[\sqrt{a S + \left(\frac{a}{2}\right)^{2} }-\frac{a}{2}\right]$=$\mathbb{E}\left[\frac{S}{\sqrt{\frac{S}{a} + \frac{1}{4} }+\frac{1}{2}}\right]$. We can see from this expression that the function $f$ is increasing in $a$ by the growth of the expectation. Thus the solution $a$ is unique.\newline
The expression of the risk-sharing rule $h_2$ as a function of $a$ is straightforward thanks to the calculations made above on the fairness condition of participant $2$. 

\begin{equation}
    \boxed{
    h_2(s)= I_2\left(\frac{J\left(s\right)}{\alpha_2}\right)= \sqrt{a s + \left(\frac{a}{2}\right)^{2} }-\frac{a}{2}.
    }
    \label{h2}
\end{equation}

\begin{equation}
    \boxed{
    h_1(s)= s - h_2(s)= s- \sqrt{a s + \left(\frac{a}{2}\right)^{2} }+\frac{a}{2}.
    }
    \label{h1}
\end{equation}

\section{Proof for the existence and uniqueness of the fixed point} 
\subsection*{Proof of Lemma \ref{lemma36}  \label{A2}}
\subsubsection*{$\varphi_{1}$ is homogeneous and monotone}
Let $\lambda>0$. If $\boldsymbol{\alpha} \ne 0$ then $(\varphi_{1}\left(\lambda\boldsymbol{\alpha}\right)(s))=J(s,\lambda\boldsymbol{\alpha})=\lambda{J(s,\boldsymbol{\alpha})}=\lambda(\varphi_{1}\left(\boldsymbol{\alpha}\right)(s))$. If $\boldsymbol{\alpha}=0$, then $(\varphi_{1}\left(\lambda\boldsymbol{\alpha}\right)(s))=\lambda(\varphi_{1}\left(\boldsymbol{\alpha}\right)(s))=0$. So the homogeneity of $\varphi_{1}$ is verified.Concerning the monotonicity, take $\alpha^{1}$ and $\alpha^{2}$ in $\mathbb{R}_{+}^{n}$ such that $\alpha^{1} \geq \alpha^{2}$. Take $s \in D$, and let $z_{1}$ and $z_{2}$ be defined by $F\left(z_{1}, \alpha^{1}\right)=s$ and $F\left(z_{2}, \alpha^{2}\right)=s$. We then have $z_{i}=\left(\varphi_{1}\left(\alpha^{i}\right)\right)(s)$ for $i=1,2$. Because the function $F(\cdot, \cdot)$ is decreasing in each of the components of its second argument and increasing in its first argument,  together with the vector inequality $\alpha^{1} \geq \alpha^{2}$ and the equality $F\left(z_{1}, \alpha^{1}\right)=F\left(z_{2}, \alpha^{2}\right)$ imply that $z_{1} \leq z_{2}$, with strict inequality as soon as $\alpha^{1}$ and $\alpha^{2}$ are not equal.

\subsubsection*{$\varphi_{1}: \mathbb{R}_{+}^{n} \rightarrow \mathcal{K}$ is continuous}
To show the continuity of $\varphi_{1}$, one makes use of the following lemma taken from \cite{pazdera_composite_2017}.
\begin{lemma}
\emph{
Let a topological space $\mathcal{Y}$ and a sequentially continuous mapping $f(\cdot, \cdot)$ from $\mathbb{R}_{+} \times \mathcal{Y}$ to $\mathbb{R}_{+}$ be given. Suppose that for every $v \in \mathcal{Y}$ there is exactly one $s \in \mathbb{R}_{+}$ such that $f(s, v)=0$. Let $\left(v_{k}\right)_{k=1,2, \ldots}$ be a sequence in $\mathcal{Y}$ that converges to $\bar{v} \in \mathcal{Y}$. Define $s_{k}$ $(k=1,2, \ldots)$ by the equations $f\left(s_{k}, v_{k}\right)=0$, and let $\bar{s}$ be defined by $f(\bar{s}, \bar{v})=0$. If the collection $\left\{s_{k} \mid k \in \mathbb{N}\right\}$ is bounded, then $\lim _{k \rightarrow \infty} s_{k}=\bar{s}$.}
\label{lemma52}
\end{lemma}

Now one can demonstrate the continuity of $\varphi_{1}$. Let $\left(\alpha^{k}\right)_{k=1,2, \ldots}$ be a sequence of vectors in $\mathbb{R}_{+}^{n}$ converging to a vector $\alpha \in \mathbb{R}_{+}^{n}$. Take $s \in A$; write $z_{k}:=\left(\varphi_{1}\left(\alpha^{k}\right)\right)(s)$ and $z=\left(\varphi_{1}(\alpha)\right)(s)$. We need to show that the sequence $\left(z_{k}\right)_{k=1,2,\ldots}$ converges to $z$.

By definition, the numbers $z_{k}$ and $z$ are positive and satisfy $F\left(z_{k}, \alpha^{k}\right)=s$ and $F(z, \alpha)=s$. Suppose there would be a subsequence $\left(z_{k_{j}}\right)_{j=1,2, \ldots}$ that tends to infinity. For all $i$ with $\alpha_{i}>0$, the sequences $\left(\alpha_{i}^{k_{j}}\right)_{j=1,2, \ldots}$ tend to finite limits,
namely $\alpha_{\mathrm{i} \cdot}$ Consequently, the quotients $z_{k_{j}} / \alpha_{i}^{k_{j}}$ tend to infinity, and therefore
$$
s=\lim _{j \rightarrow \infty} F\left(z_{k_{j}}, \alpha^{k_{j}}\right)=\sum_{i=1}^{n} \max[X_i] \text {. }
$$
However, we have $s \in [0, \sum_{i=1}^{n} \max[X_i])$ so that $s<\sum_{i=1}^{n} \max[X_i]$. From this contradiction it follows that the set $\left\{z_{k} \mid k \in \mathbb{N}\right\}$ is bounded, and it follows from Lemma \ref{lemma52} that $\lim _{k \rightarrow \infty} z_{k}=z$.

\subsection*{Proof of Lemma \ref{lemma54} \label{A5}}
Take $s \in A$. Since the entries with indices in $Q$ are assumed to be positive and those with indices in $R$ tend to infinity. we can assume that all entries of $\alpha^{k}$ are positive. Then the numbers $z_{k}=\varphi_{1}\left(\alpha^{k}\right)(s)$ are defined implicitly by
\begin{equation}
 \sum_{i \in R} I_{i}\left(z_{k} / \alpha_{i}^{k}\right)+\sum_{i \in Q} I_{1}\left(z_{k} / \alpha_{i}\right)=s \text {. }
    \label{nonsect1}   
\end{equation}
Suppose that $\left(z_{k}\right)_{k=1,2, \ldots}$ has a bounded subsequence $\left(z_{k_{j}}\right)_{j=1,2, \ldots .}$. The quotients $z_{k_{j}} / \alpha_{i}^{k_{j}}$ tend to zero for $i \in R$ so that the first term on the left-hand side in \eqref{nonsect1} tends to $0$. The quotients $z_{k_{j}} / \alpha_{i}$ for $i \in Q$ remain bounded so that the second term on the left-hand side is bounded from above. Therefore the left-hand side tends to be bounded from above which leads to a contradiction. The statement in the lemma follows.

\subsection*{Proof of Lemma \ref{lemma38} \label{A6}}
\subsubsection*{$\varphi_{2}$ is homogeneous and strictly monotone}
Let $\lambda>0$. If $J \ne 0$ then $\left(\varphi_{2}(\lambda{J})\right)_{i}=\lambda\alpha_i=\lambda\left(\varphi_{2}(J)\right)_{i}$. If $J = 0$ then $\left(\varphi_{2}(\lambda{J})\right)_{i}=\lambda\left(\varphi_{2}(J)\right)_{i}=0$. So the homogeneity of $\varphi_{2}$ is verified.\newline
Let $J_1$ and $J_2$ in $\mathcal{K}$ s.t. $J_1 > J_2$. Thanks to the actuarial fairness condition $E\left[ I_{i}\left(J_{1}(S) / \alpha_{i}^{1}\right)\right]=E\left[ I_{i}\left(J_{2}(S) / \alpha_{i}^{2}\right)\right]=E\left[X_i\right]$. As  $J_1 > J_2$ then $E\left[ I_{i}\left(J_{1}(S) / \alpha_{i}^{1}\right)\right] > E\left[ I_{i}\left(J_{2}(S) / \alpha_{i}^{1}\right)\right]$ so $E\left[ I_{i}\left(J_{2}(S) / \alpha_{i}^{2}\right)\right] > E\left[ I_{i}\left(J_{2}(S) / \alpha_{i}^{1}\right)\right].$ We can conclude that $\varphi_{2}$ is striclty monotone by observing the fact that the map $\alpha_i \mapsto E\left[I_{i}\left(J(S) / \alpha_{i}\right)\right]$ is decreasing.

\subsubsection*{$\varphi_{2}$ is sequentially continuous}
Let $\left(J_{k}\right)_{k=1,2, \ldots}$ be a sequence in $\mathcal{K}$, converging pointwise to $J \in \mathcal{K}$, and fix $i \in\{1, \ldots, n\}$. Denote $\alpha_{i}^{k}:=\left(\varphi_{2}\left(J_{k}\right)\right)_{i}$ and $\alpha_{i}:=\left(\varphi_{2}(J)\right)_{i}$. One wants to show that $\left(\alpha_{i}^{k}\right)_{k=1,2, \ldots}$ converges to $\alpha_{i}$.

First, assume that the limit function $J$ is nonzero; we can then assume that all elements of the sequence $J_{k}$ are nonzero as well. Note that the collection of random variables $J_{k}(S)$ is bounded above by $\sup_{k} \sup_{s \in A} J_{k}(s)$ and below by $\inf_{k} \inf_{s \in A} J_{k}(s)$. Hence if $\left(\alpha_{i}^{k_{j}}\right)_{j=1,2, \ldots}$ converges to infinity, one has
\begin{equation}
   E\left[X_i\right]=\lim _{j \rightarrow \infty} E\left[ I_{i}\left(J_{k_{j}}(S) / \alpha_{i}^{k_{j}}\right)\right]=0.
   \label{seq}
\end{equation}
$X_i \geq 0$ then this implies $X_i=0$ almost surely. This contradicts the assumption that $X_i$ is not equal to a constant almost surely. Consequently, the collection $\left\{\alpha_{i}^{k} \mid k \in \mathbb{N}\right\}$ is bounded. Consider the function $G: \mathbb{R}_{+} \times \mathcal{K} \rightarrow$ $\mathbb{R}_{+}$ defined by
$$
G\left(\alpha_{i}, J\right)=E\left[I_{i}\left(J(S) / \alpha_{i}\right)\right].
$$
This mapping is sequentially continuous thanks to the bounded convergence theorem. The relation $\lim \alpha_{i}^{k}=\alpha_{i}$ follows from Lemma \ref{lemma52}.

Consider now the case in which $J=0$. In this case, we have by definition $\alpha_{i}=0$. Take $\varepsilon>0$ and suppose that there would exist a subsequence $\left(\alpha_{i}^{k_{j}}\right)_{j=1,2, \ldots}$ such that $\alpha_{i}^{k_{j}}>\varepsilon$ for all $j=1,2, \ldots .$ The convergence of $\left(J_{k}\right)_{k=1,2, \ldots}$ to $J=0$ would then imply the same conclusion as in \eqref{seq}. Consequently, we have $\lim _{k \rightarrow \infty} \alpha_{i}^{k}=0$, as was to be shown.

\subsection*{Proof of Lemma \ref{lemma57} \label{A8}}
Choose $i \in\{1, \ldots, n\}$. Assume that the $i$ th entry of $\alpha^{k}:=$ $\varphi_{2}\left(J_{k}\right)$ does not tend to infinity. Then there exist a finite number $M$ and a subsequence $\left(\alpha_{i}^{k_{j}}\right)_{j=1,2, \ldots}$ such that $\alpha_{i}^{k_{j}}<M$ for all $j$. We would then have
$E\left[X_i\right]=\lim _{j \rightarrow \infty} E\left[ I_{i}\left(J^{k_{j}}(S) / \alpha_{i}^{k_{j}}\right)\right]=\max[X_i].$
This is a contradiction since it has been assumed that $E\left[X_i\right]<\max[X_i]$. Therefore the statement of the lemma follows.

\subsection*{Proof of Proposition \ref{prop1} \label{A9}}
Let $\alpha>0$ be such that $\varphi(\alpha)=\lambda \alpha$. Since $\varphi$ maps the positive cone into itself, the eigenvalue $\lambda$ must be positive. Define $J=\varphi_{1}(\alpha)$; then $\varphi_{2}(J)=\lambda \alpha$. Note that $J(s)>0$ for all $s \in A$. By definition, one has
$$
\begin{gathered}
\sum_{i=1}^{n} I_{i}(J(s) / \alpha_{i})=s \ \text{for all} \ s\ \in \ A \\
E\left[ I_{i}\left(J(S) /\left(\lambda \alpha_{i}\right)\right)\right]=E\left[X_i\right] \quad\forall \ i \ \in \{1, \dots, n \}.
\end{gathered}
$$
Therefore,
$$
\sum_{i=1}^{n} E\left[ I_{i}\left(J(S) /\left(\lambda \alpha_{i}\right)\right)\right]=\sum_{i=1}^{n} E\left[X_i\right]=E\left[S\right]=\sum_{i=1}^{n} E\left[ I_{i}\left(J(S) /\left(\alpha_{i}\right)\right)\right]
$$
The claim follows by noting that the function $\lambda \mapsto I_{i}\left(J(s) /\left(\lambda \alpha_{i}\right)\right)$, for fixed $s$ and fixed $i$, is decreasing in $\lambda$.

\subsection*{Proof of Theorem \ref{theo1} \label{A10}}
The continuity, monotonicity, and homogeneity follow from Lemmas \ref{lemma36} and \ref{lemma38}. These lemmas also imply strong monotonicity on the positive cone. Consider now two nonempty complementary subsets $Q$ and $R$ of the index set $\{1, \ldots, n\}$. If $\alpha^{1}$ and $\alpha^{2}$ are such that $\alpha_{Q}^{2}>0, \alpha_{Q}^{1}=\alpha_{Q}^{2}$, and $\alpha_{R}^{1}>\alpha_{R}^{2}$, then it follows from Lemma \ref{lemma54} that $\varphi_{1}\left(\alpha^{1}\right)>\varphi_{1}\left(\alpha^{2}\right)$. The strict inequality is preserved by the mapping $\varphi_{2}$ according to \ref{lemma38}, so that item (1) in Definition \ref{def4} is satisfied. The condition in item (2) is fulfilled due to Lemmas \ref{lemma54} and \ref{lemma57}.

\section{Proof for the convergence of the fixed point algorithm}

\subsection*{Proof of Lemma \ref{lemma44} \label{A1}}
Take $x, y \in \mathbb{R}_{++}^{\mathrm{n}}$ with $d(x, y) > 0$. Define $M := \text{max}_i(x_i/y_i)$, $m := \text{min}_i(x_i/y_i)$. We then have $my \lneq x \lneq My,$ and by homogeneity and strong monotonicity of $\varphi$ one obtains $m\varphi(y) < \varphi(x) < M\varphi(y)$. Therefore,

$$ \text{min}_i\frac{\varphi(x)_i}{\varphi(y)_i} > m, \ \ \ \text{max}_i\frac{\varphi(x)_i}{\varphi(y)_i} < M$$
and hence $d(\varphi(x), \varphi(y)) < \log(M/m) = d(x, y)$.

\subsection*{Proof of Remark \ref{prop61} \label{A14}}
Suppose that $v_1, \ldots, v_n$ are disutility functions of participants, and write $-v_i^{\prime}(s) / v_i^{\prime \prime}(s)=\sigma s+\tau_i$. By definition of the functions $I_i$ we have, for all $z>0, v_i^{\prime}\left(I_i(z)\right)=z$ and hence $v_i^{\prime \prime}\left(I_i(z)\right) I_i^{\prime}(z)=1$, so that
$$
-z I_i^{\prime}(z)=-\frac{v_i^{\prime}\left(I_i(z)\right)}{v_i^{\prime \prime}\left(I_i(z)\right)}=\sigma I_i(z)+\tau_i .
$$
For any given weight vector $\alpha \in \mathbb{R}_{++}^n$, the function $I$ defined by $I(z)=\sum_{i=1}^n I_i\left(z / \alpha_i\right)$ satisfies
$$
\begin{aligned}
-z I^{\prime}(z) &=-\sum_{i=1}^n\left(z / \alpha_i\right) I_i^{\prime}\left(z / \alpha_i\right) 
&=\sum_{i=1}^n\left(\sigma I_i\left(z / \alpha_i\right)+\tau_i\right)=\sigma I(z)+\sum_{i=1}^n \tau_i .
\end{aligned}
$$
Denote $\tau:=\sum_{i=1}^n \tau_i$. Since $J$ as defined in \eqref{form53} is the inverse function of $I$, one has
\begin{equation}
   -J(s) I^{\prime}(J(s))=\sigma s+\tau .
   \label{equ62}
\end{equation}
From the relation $I(J(s))=s$ it follows that $I^{\prime}(J(s)) J^{\prime}(s)=1$; therefore \eqref{equ62} implies that $-J(s) / J^{\prime}(s)=\sigma s+\tau$. This shows that the function $J$ defined by \eqref{form53} depends on the coefficients $\alpha_1,\dots, \alpha_n$ only through a multiplicative factor. Consequently, the coefficients $\alpha_1,\dots, \alpha_n$ that are determined from the function $J$ via \eqref{formule55} represent a positive eigenvector of $\varphi$, so that convergence of the iteration \eqref{formule58} is achieved in one step.

\section{Proof of Proposition \ref{premium} \label{pre1}}
It is known in Example \ref{toym} that the risk-sharing rules are given by:
$$   h_1(S)= S- \sqrt{a S + \left(\frac{a}{2}\right)^{2} }+\frac{a}{2}  \ \text{and} \ h_2(S)= \sqrt{a S + \left(\frac{a}{2}\right)^{2} }-\frac{a}{2},$$
    $$
       \tilde{h}_1(\tilde{S})= \tilde{S}- \sqrt{\tilde{a} \tilde{S} + \left(\frac{\tilde{a}}{2}\right)^{2} }+\frac{\tilde{a}}{2}  \ \text{and} \ \tilde{h}_2(\tilde{\tilde{S}})= \sqrt{\tilde{a} \tilde{S} + \left(\frac{\tilde{a}}{2}\right)^{2} }-\frac{\tilde{a}}{2},
    $$
where $a$ and $\tilde{a}$ are solutions of $\mathbb{E}\left[\sqrt{a S + \left(\frac{a}{2}\right)^{2} }-\frac{a}{2}\right]= \mathbb{E}\left[X_2\right]$ and $\mathbb{E}\left[\sqrt{\tilde{a} \tilde{S} + \left(\frac{\tilde{a}}{2}\right)^{2} }-\frac{\tilde{a}}{2}\right]= \mathbb{E}\left[\tilde{X}_2\right].$\newline
One can compare $a$ and $\tilde{a}$ with the hypothesis that $S \preceq_{\mathrm{CX}} \tilde{S}$. We know that $\mathbb{E}\left[h_2(S)\right]=\mathbb{E}\left[\tilde{h}_2(\tilde{S})\right]$ because $\mathbb{E}[X_2]=\mathbb{E}[\tilde{X}_2]$, which implies that 
\begin{equation}
    \mathbb{E}\left[\sqrt{a S + \left(\frac{a}{2}\right)^{2} }-\frac{a}{2}\right]=\mathbb{E}\left[\sqrt{\tilde{a} \tilde{S} + \left(\frac{\tilde{a}}{2}\right)^{2} }-\frac{\tilde{a}}{2}\right].
    \label{equ1}
\end{equation}
Since the function $h_2$ is concave, $S \preceq_{\mathrm{CX}} \tilde{S}$, and using Definition \ref{defcx}, one has
\begin{equation}
    \mathbb{E}\left[\sqrt{a S + \left(\frac{a}{2}\right)^{2} }-\frac{a}{2}\right] \geq \mathbb{E}\left[\sqrt{a \tilde{S} + \left(\frac{a}{2}\right)^{2} }-\frac{a}{2}\right].
    \label{equ2}
\end{equation}
(\ref{equ1}) and (\ref{equ2}) give the following inequality

$$
    \mathbb{E}\left[\sqrt{\tilde{a} \tilde{S} + \left(\frac{\tilde{a}}{2}\right)^{2} }-\frac{\tilde{a}}{2}\right] \geq \mathbb{E}\left[\sqrt{a \tilde{S} + \left(\frac{a}{2}\right)^{2} }-\frac{a}{2}\right].
$$
As demonstrated in the proof of Example \ref{toym}, the function $\mathbb{E}\left[\sqrt{a \tilde{S} + \left(\frac{a}{2}\right)^{2} }-\frac{a}{2}\right]$ is increasing in $a$. One concludes that
    $$
        \tilde{a}\geq a.
    $$
We note by $F_S$ and $F_{\tilde{S}}$ the distribution functions of the aggregate risks $S$ and $\tilde{S}$ respectively. To prove the result, we start by determining $F_{h_i(S)}$ as a function of $F_S$.To fix ideas, let us determine the distribution function of $h_2$ according to that of $S$. \\
Let $z \geq 0$, $\mathbb{P}\left(h_2(S) \leq z\right)= \mathbb{P}\left(\sqrt{a S + \left(\frac{a}{2}\right)^{2} }-\frac{a}{2} \leq z\right) = \mathbb{P}\left(a S \leq \left(z+\frac{a}{2}\right)^2 - \left(\frac{a}{2}\right)^{2}\right)$.\\
We deduce the following relationship
\begin{equation}
    F_{h_2(S)}(z)=F_{S}\left(\frac{z^2}{a} + z\right) \ \text{i.e.} \     f_{h_2(S)}(z)=\left(\frac{2z}{a} + 1\right)f_{S}\left(\frac{z^2}{a} + z\right).
   \label{frdh2}
\end{equation}
The same method leads to 
\begin{equation}
    F_{h_1(S)}(z)=F_{S}\left(z + a \sqrt{z}\right) \ \text{i.e.} \ f_{h_1(S)}(z)=\left(\frac{a}{2\sqrt{z}} + 1\right)f_{S}\left(z + a \sqrt{z}\right).
   \label{frdh1}
\end{equation}
Our aim is to show that for all $i \ \in \ \{1, 2\}.$
$$h_i(S) \preceq_{\mathrm{CX}} \tilde{h}_i(\tilde{S})$$
Since $S \ \sim \ \Gamma(\mu, \lambda)$ then $f_{S}(s)=\frac{\lambda^{\mu}}{\Gamma(\mu)} s^{\mu - 1} \exp{(-\lambda s)}$. Equation \eqref{frdh2} leads to
$$
       f_{h_2(S)}(z)=\left(\frac{2z}{a} + 1\right)\frac{\lambda^{\mu}}{\Gamma(\mu)} \left(\frac{z^2}{a} + z\right)^{\mu - 1} \exp{\left(-\lambda \left(\frac{z^2}{a} + z\right)\right)}. 
$$
Our goal is to compare the random variables $h_2(S)$ and $\tilde{h}_2(\tilde{S})$ under the convex order. Notice that both random variables are continuous and have support on the open interval $(0,\infty)$. Define the transform $\rho_X(z)=\frac{\mathrm{d}}{\mathrm{d} z} \ln f_X(z)$ and let $\gamma(z)=\rho_{\tilde{h}_2(\tilde{S})}(z)-\rho_{h_2(S)}(z)$ for $z \in(0,\infty)$. A sufficient condition for $h_2(S) \preceq_{\mathrm{ICX}} \tilde{h}_2(\tilde{S})$ ($\mathrm{ICX}$ refers to the increasing convex order) from \cite{hesselager_order_1995} is that there exists a $c$ such that $\gamma(z)$ is negative for $z \in(0, c)$ and positive for $z \in(c, \infty)$. We have that
\begin{equation}
   \rho_{h_2(S)}(z)= \frac{1}{z+\frac{a}{2}} + \left( \frac{\mu -1}{z+\frac{z^2}{a}}- \lambda\right)\left(1+\frac{2z}{a}\right). 
\end{equation}
One may verify that $\lim _{z \rightarrow 0^{+}} \gamma(z)=-\infty$ and $\lim _{z \rightarrow \infty} \gamma(z)=\infty$. One may further verify that $\frac{\mathrm{d}}{\mathrm{d} z} \gamma(z)$ is positive for $z \in(0,\infty)$. It follows that $\gamma(z)$ satisfies the sufficient condition of \cite{hesselager_order_1995} and we have $h_2(S) \preceq_{ICX} \tilde{h}_2(\tilde{S})$. Since $\mathbb{E}\left[h_2(S)\right]=E\left[\tilde{h}_2(\tilde{S})\right]$, we also have (see Theorem 1.5.3 of \cite{muller_comparison_2002}) that $h_2(S) \preceq_{CX} \tilde{h}_2(\tilde{S})$, as required.

\bibliographystyle{apalike}
\bibliography{bibliography}
\end{document}